%% file: main.tex
\documentclass[journal]{IEEEtran}

\usepackage{amsmath,amssymb,amsfonts}
\usepackage{algorithmic}
\usepackage{algorithm}
\usepackage{array}
\usepackage{balance}
\usepackage{cases}
\usepackage{caption}   
\usepackage{cite,hyperref}
\usepackage[capitalise]{cleveref}
\usepackage{color}
\usepackage{graphicx}
\usepackage{mathtools}
\usepackage{stfloats}
\usepackage{subcaption}
\usepackage{tcolorbox}
\usepackage{textcomp}
\usepackage{theorem}
\usepackage{tikz}
\usepackage{url}
\usepackage{verbatim}
\usepackage{xcolor}

\usetikzlibrary{shapes,arrows}
\input{styles/mysymbol.sty}

\usepackage{moresize}
\usepackage{pgfplots}
\pgfplotsset{compat=1.17}
\pgfplotstableset{col sep=comma}
\input{hidden_nodes_defs}

\newtcolorbox{myblockt}[1]{colback=urblue!5!white,
	colframe=urblue,fonttitle=\bfseries,
	title=#1}
\newtcolorbox{myblock}{colback=urblue!5!white,
	colframe=urblue,fonttitle=\bfseries}

\begin{document}

%%%%%%%%%% Tikz and pgfplots settings %%%%%%%%%%
\tikzset{every mark/.append style={scale=1.5, solid}, font=\footnotesize}
\pgfplotsset{
    width=1.05\textwidth,
    %tick label style={font=\footnotesize},
    %label style={font=\footnotesize},
    legend style={
        font=\ssmall ,  %\scriptsize,  %\ssmall,
        inner xsep=1pt,
        inner ysep=1pt,
        nodes={inner sep=1pt}},
    legend cell align=left,
	every axis/.append style={line width=0.5pt},
	every axis plot/.append style={line width=1.25pt},
    every axis y label/.append style={yshift=-5pt}
}

\title{Joint Network Topology Inference in the Presence of Hidden Nodes}

\author{Madeline Navarro,~\IEEEmembership{Student Member,~IEEE}, Samuel Rey,~\IEEEmembership{Member,~IEEE}, Andrei Buciulea,~\IEEEmembership{Student Member,~IEEE}, Antonio G. Marques,~\IEEEmembership{Senior Member,~IEEE}, and Santiago Segarra,~\IEEEmembership{Senior Member,~IEEE}% <-this % stops a space
\thanks{
This work was partially supported by NSF under award CCF-2008555, Spanish AEI Grants FPU17-04520,  PID2019-105032GB-I00, PID2022-136887NB-I00, and by the Autonomous Community of 
Madrid within the ELLIS Unit Madrid framework and URJC grant F861.
Research was sponsored by the Army Research Office and was accomplished under Grant Number W911NF-17-S-0002. The views and conclusions contained in this document are those of the authors and should not be interpreted as representing the official policies, either expressed or implied, of the Army Research Office or the U.S. Army or the U.S. Government. The U.S. Government is authorized to reproduce and distribute reprints for Government purposes notwithstanding any copyright notation herein.
Preliminary results were presented at ICASSP 2022 \cite{rey2022joint}.
\emph{(Corresponding author: M. Navarro)}

M. Navarro and S. Segarra are with the Dept. of ECE, Rice University, Houston, TX 77005 USA (e-mail: \{nav,segarra\}@rice.edu). 
S. Rey, A. Buciulea, and A. G. Marques are with the Dept. of Signal Theory and Comms., King Juan Carlos University, 28933 Madrid, Spain (e-mail: \{samuel.rey.escudero, andrei.buciulea, antonio.garcia.marques\}@urjc.es).}%
}

\maketitle
%%%%%%%%%%%%%%%%%%%%%%%%%%%%%%%%%%%%%%%%%%%%%%%%%%%%%%%%%%%%%%%

%%%%%%%%%%%%%%%%%%%%%%%%%%%%%%%%%%%%%%%%%%%%%%%%%%%%%%%%%%%%%%%
\begin{abstract}
    We investigate the increasingly prominent task of jointly inferring \emph{multiple} networks from nodal observations.
    While most \emph{joint} inference methods assume that observations are available at all nodes, we consider the realistic and more difficult scenario where a subset of nodes are \emph{hidden} and cannot be measured.
    Under the assumptions that the partially observed nodal signals are graph stationary and the networks have similar connectivity patterns, we derive structural characteristics of the connectivity between hidden and observed nodes.
    This allows us to formulate an optimization problem for estimating networks while accounting for the influence of hidden nodes.
    We identify conditions under which a convex relaxation yields the sparsest solution, and we formalize the performance of our proposed optimization problem with respect to the effect of the hidden nodes.
    Finally, synthetic and real-world simulations provide evaluations of our method in comparison with other baselines.
\end{abstract}

%%%%%%%%%%%%%%%%%%%%%%%%%%%%%%%%%%%%%%%%%%%%%%%%%%%%%%%%%%%%%%%

\section{Introduction}\label{S:intro}
% Why graphs
% \IEEEPARstart{I}{n} recent years, graphs have become a crucial tool for modeling the irregular (non-Euclidean) structure commonly found in contemporary data.
% This strategy has enjoyed success in a range of applications such as genetics, brain networks, and communications, where disciplines like machine learning or signal processing rely on graphs to capture the underlying irregular domain of the signals~\cite{kolaczyk2009book,sporns2012book,ortega2018graph}.
\IEEEPARstart{I}{n} recent years, graphs have become a staple model of the irregular (non-Euclidean) structure commonly found in contemporary data.
Disciplines like signal processing often rely on graphs to capture the underlying irregular domain of the signals, where such successful applications include genetics, brain networks, and communications~\cite{kolaczyk2009book,sporns2012book,ortega2018graph}.
Nevertheless, despite the popularity of graph-based methods, in practice the topology of the graph is often not readily available, spurring the development of graph learning algorithms~\cite{friedman2008sparse,mateos2019connecting,xia2021graph} to infer the network topology from a set of nodal observations.

% Why NTI and Joint NTI
Indeed, the task of \emph{network topology inference}, also known as \emph{graph learning}, has emerged as a vibrant research area within graph signal processing (GSP)~\cite{shuman2013emerging,sandryhaila2013discrete,djuric2018cooperative,rey2022robust}.
% , a field devoted to developing tools for processing graph signals
A crucial assumption for learning the graph topology is the statistical relationship between the signals and the unknown topology.
% that the (statistical) properties of the signals depend on the unknown topology.
Different assumptions lead to different methods, with noteworthy examples including correlation networks and (Gaussian) Markov random fields ((G)MRF)~\cite{meinshausen06,friedman2008sparse,kolaczyk2009book}, smooth (local total variation) models~\cite{kalofolias2016learn,dong2016learning,saboksayr2021accelerated}, GSP-based approaches~\cite{segarra2017network,egilmez2017graph,shafipour2020online}, and models with more elaborate graph priors~\cite{roddenberry2021network,rey2022enhanced}.
A common feature of the previous works is that they focus on learning a single graph.
However, many contemporary setups involve \emph{multiple related networks}, each with a subset of signals.
 % or can be described by \emph{multilayered-graphs}
Some examples include brain analytics, where observations from different \textit{patients} are used to estimate their brain functional networks; social networks, where the same set of users may present different types of \textit{interactions}; or multi-hop communication networks in dynamic environments, where a network needs to be inferred for each \textit{time instant}.
% Intuitively, in situations where several closely related networks exist, approaching the problem in a joint fashion can boost the performance of network topology inference by enabling us to harness the relationships among graphs~\cite{murase2014multilayer,danaher2014joint,navarro2022joint,arroyo2021inference,navarro2022jointb, wang2020high}.
Intuitively, in situations where several closely related networks exist, approaching the problem in a joint fashion can boost the performance of network topology inference by harnessing the relationships among graphs~\cite{murase2014multilayer,danaher2014joint,navarro2022joint,arroyo2021inference,navarro2022jointb, wang2020high}.

% Challenges of hidden nodes
Despite the clear benefits, joint network topology inference approaches usually assume that observations from every node are available, which is often not the case.
In many relevant scenarios, the observed signals correspond only to a subset of the nodes in the whole graph, while the remaining nodes stay unobserved or \emph{hidden}.
Ignoring the presence of the hidden nodes can drastically hinder the performance of the graph learning algorithms.
Nevertheless, accounting for their influence is not a trivial endeavor since the inference task becomes ill-posed.
For \textit{single network} inference,
%In the context of learning a single graph, 
some works dealing with this challenging setting include graphical models~\cite{chandrasekaran2012latent,chang2019graphical}, inference of linear Bayesian networks~\cite{anandkumar2013learning}, nonlinear regression \cite{mei2018silvar}, and stationary-based algorithms~\cite{buciulea2019network,buciulea2022learning}.
However, the presence of hidden nodes is yet to be addressed for several unknown graphs.
% Recall that the cornerstone of joint topology inference is to harness the similarity of the graphs, and hence, modeling the influence of hidden nodes becomes even more critical since it is unclear how to measure the graph similarity between nodes that remain unobserved.
Since the key to joint topology inference is exploiting the similarity of the graphs, it is crucial to model the influence of the hidden nodes to measure the graph similarity between nodes that remain unobserved.

% Our solution
To this end, we propose a topology inference method that simultaneously performs \emph{joint estimation of multiple graphs} and \emph{accounts for the presence of hidden variables}.
% To address the aforementioned limitation, we propose a topology inference method that simultaneously performs \emph{joint estimation of multiple graphs} and \emph{accounts for the presence of hidden variables}.
Under the assumption that the observed signals are realizations of a random process that is \emph{stationary} on the graph \cite{djuric2018cooperative,marques2017stationary}, we formalize the relationship between the nodal observations and the unknown networks under the influence of the hidden nodes.
The joint formulation necessitates exploiting graph similarities, not only with respect to observed nodes but also to hidden ones.
To accomplish this, we carefully model the structure associated with latent variables and exploit it with a regularization inspired by the group Lasso penalty~\cite{simon2013sparse}.
% To accomplish this, we carefully model the structure inherent to the presence of latent variables and exploit it with a regularization inspired by the group Lasso penalty~\cite{simon2013sparse}.
Finally, we conduct thorough mathematical and numerical analyses of the proposed approach, where we show the conditions under which it recovers the sparsest solution and bounds the error of the estimated graphs, and we evaluate its performance and the hidden variables' detrimental influence through simulations with synthetic and real-world data.
% After introducing the proposed approach, we conduct a thorough mathematical analysis, which shows the conditions under which we recover the sparsest solution and bound the error of the estimated graphs, and numerical analysis evaluating its performance.
% After introducing the proposed approach, we conduct a thorough mathematical analysis and show that, under certain conditions, we recover the sparsest solution and bound the error of the estimated graphs.
% Finally, we evaluate the proposed method's performance and the hidden variables' detrimental influence through numerical analysis with synthetic and real-world data.

\vspace{1mm}
\noindent 
{\bf Related work and contributions.}
Early methods for joint graph learning were introduced in~\cite{danaher2014joint} assuming that observations follow a GMRF and, later on, in~\cite{navarro2022joint} followed by a joint inference method for graph stationary signals.
However, both works assumed that observations from the whole graphs were available.
At the same time, the influence of hidden nodes when learning a single graph was studied in~\cite{chandrasekaran2012latent} and~\cite{buciulea2022learning} assuming that the observations adhered respectively to a GMRF or a graph-stationary model.
On the other hand, the relevant task of learning several graphs in the presence of hidden nodes has only been considered under GMRF assumptions in the preliminary results from~\cite{rey2022jointb}. 
In contrast, in this paper, we (i) build upon our previous work from~\cite{rey2022joint} for joint graph learning with hidden variables under the more lenient assumption of stationary observations; and (ii) develop a theoretical analysis to characterize how the hidden nodes influence the quality of the estimated graphs.
Finally, note that GMRF and graph stationarity are intrinsically different models for the observations, resulting in materially different inference algorithms and, even more relevant for the problem at hand, requiring different methods to encourage graph similarities with respect to both observed and hidden nodes.

% Summary of contributions
To summarize, our main contributions are:
\begin{itemize}
    \item We design a convex optimization problem to jointly learn the topology of several related graphs in the presence of hidden variables under graph-stationary observations. 
    %Proposed a regularization to exploit the graph similarity among nodes that are not being observed.
    \item We rely on a regularization inspired by group Lasso to model the similarity between hidden nodes and hence harness the similarity of the entire node set, both hidden and observed nodes.
    \item We derive theoretical guarantees for the recoverability of the estimated graphs in the presence of hidden nodes.
    % \item We evaluate the performance of the proposed approach and the hidden variables' detrimental influence through numerical analysis and compare our approach with state-of-the-art alternatives in synthetic and real-world datasets.
    \item We evaluate the performance of the proposed approach and compare it with state-of-the-art alternatives in synthetic and real-world datasets.
\end{itemize}

% Outline
The remainder of the paper is organized as follows.
\cref{S:background} introduces GSP concepts necessary for our proposed network topology inference method and its theoretical guarantees. 
We introduce in \cref{S:problem} the task of learning graphs in the presence of hidden nodes.
In \cref{S:method} we present our proposed optimization problem that accounts for hidden nodes, along with its convex relaxation.
We provide theoretical guarantees for the viability and performance of our method in \cref{S:theory}, which are validated by several synthetic and real-world experiments in \cref{S:results}.
Finally, a concluding discussion is provided in \cref{S:conclusion}.

\section{Fundamentals of GSP}\label{S:background}

We introduce notation and concepts in GSP to characterize the statistical relationship between the network topology and measurements on nodes, both observed and hidden.

\vspace{1mm}
\noindent{\bf Notation.}
For a matrix $\bbY\in\mathbb{R}^{M\times N}$, $\mathrm{vec}(\bbY)\in\mathbb{R}^{MN}$ denotes the vertical concatenation of the columns of $\bbY$.
We let calligraphic letters denote index sets, where, given any matrix $\bbX\in\mathbb{R}^{M\times N}$ and any vector $\bbx\in\mathbb{R}^N$, we let $\bbX_{\ccalC,\cdot}$ and $\bbX_{\cdot,\ccalC}$ respectively return the rows and columns of $\bbX$ selected from index set $\ccalC$ and $\bbx_{\ccalC}$ returns the entries of $\bbx$ selected from $\ccalC$.
The notation $\bbI_M$ denotes the identity matrix of size $M\times M$, while ${\bf 1}_{M\times N}$ and ${\bf 0}_{M\times N}$ respectively represent matrices of all ones and zeros of size $M\times N$.
We let $\ccalD$, $\ccalL$, and $\ccalU$ respectively denote the indices of the diagonal, lower triangular, and upper triangular entries of a vectorized square matrix, i.e., for any matrix $\bbY\in\mathbb{R}^{M\times M}$ and $\bby = \mathrm{vec}(\bbY)$, we have that $\bby_{\ccalD}$ contains the diagonal entries of $\bbY$.
We define $\bby_{\ccalL}$ and $\bby_{\ccalU}$ similarly.
The notation $O(\cdot)$ and $o(\cdot)$ denote the usual asymptotic meaning, and we say that $f \asymp g$ if $f = O(g)$ and $g = O(f)$.

\vspace{1mm}
\noindent{\bf Graph signal processing and graph stationarity.}
We consider undirected graphs of the form $\ccalG=(\ccalV,\ccalE)$, where $\ccalV$ denotes the set of $|\ccalV| = N$ nodes and $\ccalE\subseteq\ccalV\times\ccalV$ is the edge set such that the unordered pair $(i,j)\in\ccalE$ if and only if nodes $i$ and $j$ are connected.
A convenient representation for the structure of a graph is its weighted adjacency matrix $\bbA\in\mathbb{R}^{N\times N}$, where $A_{ij}=A_{ji}\neq 0$ if and only if $(i,j)\in\ccalE$.
We may define a more general class of matrices to encode graph structure known as the graph shift operator (GSO), of which the adjacency matrix is an example~\cite{djuric2018cooperative,sandryhaila2013discrete,shuman2013emerging}.
Formally, the GSO is a square matrix $\bbS\in\mathbb{R}^{N\times N}$, where $S_{ij}\neq0$ only if $i=j$ or $(i,j)\in\ccalE$.
When $\ccalG$ corresponds to an undirected graph, the GSO $\bbS$ is symmetric, where $S_{ij} = S_{ji}$ are assigned the same value associated with the edge $(i,j)$.
Commonly chosen GSOs include the adjacency matrix $\bbA$ and the graph Laplacian $\bbL := \mathrm{diag}(\bbA{\bf 1}) - \bbA$~\cite{djuric2018cooperative,shuman2013emerging}.
Because we consider undirected graphs, $\bbS$ is symmetric and thus diagonalizable.

% \medskip
% \noindent\textbf{Graph filters.}
Critical to the network inference task is the statistical relationship between nodal observations and the topology of $\ccalG$.
We represent real-valued observations on the nodes of $\ccalG$ as graph signals $\bbx=[x_1,\dots,x_N]^\top\in\mathbb{R}^N$, where $x_i$ denotes the signal value at the $i$-th node.
In this work, we assume that the \emph{observations} are realizations of a random graph signal that is \emph{stationary on the GSO $\bbS$ associated with graph} $\ccalG$ ~\cite{marques2017stationary,segarra2017network,pasdeloup2017characterization}, a versatile model that has shown theoretical and practical relevance. From a mathematical point of view, a random graph signal $\bbx$ is \emph{stationary} on a GSO $\bbS$ if the covariance matrix of $\bbx$, denoted as $\bbC$, can be written as a (matrix) polynomial of the GSO $\bbS$, which results in  $\bbC$ and  $\bbS$ having the same eigenvectors~\cite{djuric2018cooperative,perraudin2017stationary,girault2015translation,marques2017stationary}.  This definition includes correlation networks, where $\bbC=\bbS$ and MRFs, where $\bbC=\bbS^{-1}$, as particular cases. From a practical (generative) point of view,  stationary random graph signals are particularly suited to  represent consensus dynamics, heat diffusion processes, and network processes on brain structural networks~\cite{zhu2020network,thanou2017learning,li2019identifying}.
% 
% \red{Formally, we model the random graph signal $\bbx$ as $\bbx=\bbH\bbw$, where $\bbw$ is a stochastic zero-mean white input signal and $\bbH$ has the form $\bbH=\sum_{l=0}^{L-1} h_l \bbS^l$ with real-valued filter coefficients $\{h_l\}_{l=0}^{L-1}$ that diffuses the signal $\bbw$ according to the network structure encoded in $\bbS$~\cite{sandryhaila2013discrete,segarra2017optimal,isufi2022graphfiltersurvey}.}
% \red{The diffusion model sufficiently models nodal behavior for many signal processing tasks, including denoising and interpolation~\cite{djuric2018cooperative,segarra2017optimal,segarra2017blind,zhu2020estimating,zhu2020network}}.
% 
Formally, under this point of view we have that the random graph signal $\bbx$ can be modelled as $\bbx = \bbH\bbw$, where $\bbw$ is a stochastic zero-mean white input signal and $\bbH$ performs the diffusion process on $\bbw$ that characterizes the influence of the GSO $\bbS$ on $\bbx$. 
To that end, the matrix $\bbH$ is assumed to be a \emph{linear graph filter}~\cite{sandryhaila2013discrete,segarra2017optimal,isufi2022graphfiltersurvey}, a matrix polynomial of the GSO $\bbH = \sum_{l=0}^{L-1} h_l\bbS^l$ with order $L$ and real-valued filter coefficients $\{h_l\}_{l=0}^{L-1}$ that sufficiently models nodal behavior for many signal processing tasks, including denoising and interpolation~\cite{djuric2018cooperative,segarra2017optimal,segarra2017blind,zhu2020estimating,zhu2020network}.
The structure of $\bbS$ dictates the behavior of the graph signal $\bbx=\bbH\bbw$, where we may view $\bbS^l \bbw$ as the diffusion of $\bbw$ across an $l$-hop neighborhood.
Under the diffusion model, the signal behavior at the $i$-th node is encoded in the diffused signal values in an $(L-1)$-hop radius.
Under this setting, the graph signals are \emph{random} with covariance $\bbC = \mathbb{E}[\bbx\bbx^\top]=\bbH\mathbb{E}[\bbw\bbw^\top]\bbH^\top=\bbH\bbH^\top=\bbH^2$ due to the input $\bbw$ being white.
Clearly, if $\bbH$ is a polynomial of $\bbS$, so is $\bbC=\bbH^2$, showing that both points of view are equivalent.

%We call a graph signal \emph{stationary} on its underlying graph $\ccalG$ if the graph signal covariance matrix $\bbC$ and the GSO $\bbS$ are both diagonalized by the same eigenvectors~\cite{djuric2018cooperative,perraudin2017stationary,girault2015translation,marques2017stationary}. Notice that the definition $\bbx=\bbH\bbw$ for a GSO polynomial $\bbH$ and a zero-mean white vector $\bbw$ implies that the graph signal $\bbx$ is stationary.  
Finally, we note that under stationarity of $\bbx$, we have that matrices $\bbS$ and $\bbC$ commute and hence, it must hold that $\bbC\bbS=\bbS\bbC$. 
This is a compact and tractable way to account for the graph stationarity of the observed signals and will be later on used as a constraint in our optimization problems.

\section{Inference of multilayered graphs with latent variables}\label{S:problem}

Let there be a set of $K$ undirected networks $\{\ccalG^{(k)}\}_{k=1}^K$ on the same set $\ccalV$ of $N$ nodes with GSOs denoted as $\{\bbS^{*(k)}\}_{k=1}^K$.
We assume that for each graph there exist a set with $R_k$ realizations of a \emph{stationary} graph signal collected in data matrices $\bbX^{(k)}\in\mathbb{R}^{N\times R_k}$, where the $R_k$ columns contain the nodal observations on the $k$-th graph.
For a signal $\bbx^{(k)}$ on the $k$-th graph, its covariance matrix is denoted by $\bbC^{(k)}=\mathbb{E}[\bbx^{(k)}(\bbx^{(k)})^\top]$.
We further assume that for every graph we do not know the entire data matrix $\bbX^{(k)}$ but only observe signal values on a subset $\ccalO\subset\ccalV$ of $O$ nodes, where $\ccalH:=\ccalV\backslash\ccalO$ denotes the set of $H$ hidden nodes.
Our goal is to \emph{estimate the subnetwork of each network $\ccalG^{(k)}$ induced by $\ccalO$ from partially observed graph signals}.

Under this setting, we can now formalize the task of estimating the network structure at the node subset $\ccalO$ that is encoded in the GSOs $\{\bbS^{*(k)}\}_{k=1}^K$.
Without loss of generality, we partition the GSO and the covariance matrix of each network as
\begin{equation}\label{e:mat_partition}
    \bbS^{*(k)}=
    \left[\begin{matrix}
    \bbSo^{*(k)} & \bbSoh^{*(k)} \\
    \bbSho^{*(k)} & \bbSh^{*(k)}
    \end{matrix}\right],
    \quad
    \bbC^{(k)}=
    \left[\begin{matrix}
   \bbCo^{(k)} & \bbCoh^{(k)} \\
    \bbCho^{(k)} & \bbCh^{(k)}
    \end{matrix}\right],
\end{equation}
where $\bbSoh^{*(k)}=(\bbSho^{*(k)})^\top$ and $\bbCoh^{(k)}=(\bbCho^{(k)})^\top$ by the symmetry of $\bbS^{*(k)}$ and $\bbC^{(k)}$.
The submatrices $\bbSo^{*(k)}\in\mathbb{R}^{O\times O}$ and $\bbSh^{*(k)}\in\mathbb{R}^{H\times H}$ encode the connectivity of the subnetworks of $\ccalG^{(k)}$ induced by $\ccalO$ and $\ccalH$, respectively, while $\bbSoh^{*(k)}\in\mathbb{R}^{O\times H}$ represents the edges connecting observed nodes to hidden nodes. 
We similarly define $\bbCo^{(k)}$, $\bbCh^{(k)}$, and $\bbCoh^{(k)}$.
Given the partitions in \eqref{e:mat_partition}, we aim to estimate the subnetworks encoded in $\{\bbSo^{*(k)}\}_{k=1}^K$.

We also partition each $\bbX^{(k)}$ to be conformal with $\bbS^{*(k)}$ and $\bbC^{(k)}$ as $\bbX^{(k)}=[\bbXo^{(k)\top},\bbXh^{(k)\top}]^\top$, where $\bbXo^{(k)}\in\mathbb{R}^{O\times R_k}$ is the data matrix containing the partially observed graph signals and $\bbXh^{(k)}\in\mathbb{R}^{H\times R_k}$ remains unknown.
We can thus apply the partially observed \emph{stationary} graph signals $\bbXo^{(k)}$ and the commutative relationship $\bbC^{(k)}\bbS^{*(k)}=\bbS^{*(k)}\bbC^{(k)}$ as described in Section~\ref{S:background} to recover the structure in $\bbSo^{*(k)}$.
% While we cannot compute the entire sample covariance matrices $\hat{\bbC}^{(k)}$, we may obtain the submatrices $\hbCo^{(k)} = \frac{1}{R_k}\bbXo^{(k)}(\bbXo^{(k)})^\top$ 
% \red{SS: Do we need this sentence about the sample covariance here? Maybe there is a better place for it since we do not use the sample covariance in all this section?}. 
Given the problem setting, we can now formalize our joint topology inference problem in the presence of hidden nodes as follows.

\medskip

\noindent{\bf Problem 1}\emph{
Given the sets $\{\bbXo^{(k)}\}_{k=1}^K$ of graph signal values at the observed nodes for each of the $K$ graphs, recover the sparsest $\{\bbSo^{*(k)}\}_{k=1}^K$ under the following assumptions: \\
(AS1) the number of hidden nodes $H$ is much smaller than the number of observed nodes, that is, $H\ll O$; \\
(AS2) the signals in $\bbX^{(k)}$ are realizations of a process that is stationary in $\bbS^{*(k)}$; and \\
(AS3) the GSOs $\bbS^{*(k)}$ and $\bbS^{*(k')}$ for $k\neq k'$ are sparse and have similar sparsity patterns, that is, $\bbS^{*(k)}-\bbS^{*(k')}$ is sparse.
}

% \noindent{\bf Problem 1}\emph{
% Given the sets, $\{\bbXo^{(k)}\}_{k=1}^K$ of graph signal values at the observed nodes for each of the $K$ graphs, find the sparsest matrices $\{\hbSo^{(k)}\}_{k=1}^K$ that encode the network structure at the observed nodes under the following assumptions: \\
% (AS1) the number of hidden nodes $H$ is much smaller than the number of observed nodes, that is, $H\ll O$; \\
% (AS2) the signals in $\bbX^{(k)}$ are realizations of a process that is stationary in $\bbS^{*(k)}$; and \\
% (AS3) the GSOs $\bbS^{*(k)}$ and $\bbS^{*(k')}$ have similar sparsity patterns.
% }

\medskip

We elaborate on the implications of the assumptions.
The first assumption \textit{(AS1)} ensures the tractability of the problem.
When most of the nodes in the graph are observed, the covariance submatrix $\bbCo^{(k)}$ sufficiently characterizes the structure of $\bbSo^{*(k)}$.
Importantly, under $H\ll O$, the matrix product $\bbCoh^{(k)}\bbSho^{*(k)}$ is low-rank, a crucial result for inferring $\bbSo^{*(k)}$, which is also assumed in different single graph-learning approaches.
Assumption \textit{(AS2)} establishes a global relationship between the graph signals $\bbX^{(k)}$ and the unknown graph structure $\bbS^{*(k)}$, including both observed and hidden nodes.
This assumption enables us to specify how the hidden nodes affect $\bbX^{(k)}$ by considering the connectivity between observed and hidden nodes encoded in $\bbSoh^{*(k)}$ from \eqref{e:mat_partition} and the commutative relationship $\bbC^{(k)}\bbS^{*(k)}=\bbS^{*(k)}\bbC^{(k)}$.
The final assumption guarantees that all $K$ graphs have similar edge connectivity patterns across all the shared node set $\ccalV$.
Not only can we then benefit from jointly inferring the observed subnetworks, but we may also share hidden node information across all $K$ graphs during inference.
We naturally expect that the support of $\bbSo^{*(k)}$ will be similar across all $K$ graphs~\cite{rey2022jointb,navarro2022joint,danaher2014joint}; however, it is important to also exploit the edgewise similarity for $\bbSoh^{*(k)}$ to account for connections between observed and hidden nodes.
As multiple matrices $\bbS^{(k)}$ satisfy $\bbC^{(k)}\bbS^{(k)}=\bbS^{(k)}\bbC^{(k)}$, we require additional structural priors on our target GSOs.
Thus, we select as our target GSOs $\{\bbS^{*(k)}\}_{k=1}^K$ the sparsest ones that satisfy the commutativity assumption, resulting in parsimonious network representations that are interpretable and computationally friendly.

Notice that for the simpler case where the set $\ccalH$ of hidden nodes differs across graphs, 
\textit{(AS3)} would allow us to exploit nodal observations from graph $k$ that are hidden for graph $k'$ to account for hidden nodes.
However, in this work, we address the more challenging scenario in Problem 1, where there is a subset of nodes for which there are no direct observations for \emph{any} graph.
We rely on the statistical relationship between the graph signals and the graph topology to formulate a suitable optimization problem jointly infer the subnetworks in $\bbSo^{*(k)}$.

\section{Joint graph learning with latent variables as a convex optimization problem }\label{S:method}

Network topology inference with stationary graph signals commonly exploits the commutativity of the graph signal covariance matrices and the GSOs.
We also adopt this approach; however, unlike previous works, we cannot directly apply the commutative relationship due to the presence of hidden nodes.
We must revisit the commutativity of $\bbC^{(k)}$ and $\bbS^{*(k)}$ with the partitions in \eqref{e:mat_partition} before introducing our inference problem with stationary graph signals.
From stationarity \textit{(AS2)}, we know that $\bbS^{*(k)}\bbC^{(k)}=\bbC^{(k)}\bbS^{*(k)}$ for all $k=1,\dots, K$.
From \eqref{e:mat_partition} it then follows that 
\begin{equation}\label{e:commut}
   \bbCo^{(k)}\bbSo^{*(k)}
    - \bbSo^{*(k)}\bbCo^{(k)}
    =
    (\bbP^{*(k)})^\top
    - \bbP^{*(k)}
\end{equation}
% \red{SS: Given the definition of $\bbP^{*(k)}$, should the RHS of the above equation be $(\bbP^{*(k)})^\top - \bbP^{*(k)}$ instead?}
for all $k=1,\dots,K$, where $\bbP^{*(k)} := \bbCoh^{(k)}\bbSho^{*(k)}$.
The right-hand side of \eqref{e:commut} fully accounts for the influence of hidden nodes.
When $\bbP^{*(k)}$ is known, estimating $\bbSo^{*(k)}$ relies solely on the commutator on the left-hand side.
This is similar to traditional network inference with stationary graph signals, where we also know the value of the commutator $\bbC^{(k)}\bbS^{*(k)}-\bbS^{*(k)}\bbC^{(k)}={\bf 0}_{N\times N}$.

With the prior structural information in place, we can approach estimating the subnetworks from sample covariance submatrices $\hbCo^{(k)} = \frac{1}{R_k}\bbXo^{(k)}(\bbXo^{(k)})^\top$ by the following nonconvex optimization problem
\begin{alignat}{3}&
    \min_{ \{\bbSo^{(k)},\bbP^{(k)}\}_{k=1}^K } ~
    \sum_{k=1}^K \alpha_k \| \bbSo^{(k)} \|_0
     + \sum_{k<k'} \beta_{k,k'} \| \bbSo^{(k)}-\bbSo^{(k')} \|_0
    &\nonumber\\&
    \qquad \qquad \qquad 
     + \sum_{k=1}^K \gamma_k \| {\bbP}^{(k)} \|_{2,1}
     + \sum_{k<k'} \eta_{k,k'} \left\| \left[\begin{matrix}\bbP^{(k)}\\\bbP^{(k')}\end{matrix}\right] \right\|_{2,1}
    &\nonumber\\&
    \st ~\!\!
    \textstyle\sum_{k=1}^K \| \hbCo^{(k)}\bbSo^{(k)} - \bbSo^{(k)}\hbCo^{(k)} + \bbP^{(k)} - (\bbP^{(k)})^\top \|_F^2 \leq \epsilon^2,
    &\nonumber\\&
    \qquad ~\!\!
    \bbSo^{(k)}\in\ccalS,
    \label{e:l0_prob}
&\end{alignat}
where we have introduced auxiliary matrices $\{\bbP^{(k)}\}_{k=1}^K$ to account for the right hand side of \eqref{e:commut}.
We first discuss \eqref{e:l0_prob} as it relates to the GSO submatrix estimates $\{\bbSo^{(k)}\}_{k=1}^K$.
The first two terms in the objective of \eqref{e:l0_prob} encourage sparse subnetworks with similar sparsity patterns as in \textit{(AS3)}.
The second constraint encourages valid GSOs for $\bbSo^{(k)}$.
In this work, we let the GSOs denote adjacency matrices, so we define
\begin{equation}\label{e:valid_adjmat}
    \ccalS := \left\{\bbS~:~\bbS=\bbS^\top,~\mathrm{diag}(\bbS)={\bf 0},~\textstyle\sum_j\bbS_{j1}=1\right\},
\end{equation}
where $\{\bbSo^{(k)}\}_{k=1}^K$ denote valid submatrices of nontrivial adjacency matrices, that is, $\bbSo^{(k)}\neq{\bf 0}_{O\times O}$.
While we select adjacency matrices as GSOs, problem \eqref{e:l0_prob} accommodates other GSOs, such as the graph Laplacian~\cite{segarra2017network}, under minor modifications.

We next discuss the estimated auxiliary matrices $\{\bbP^{(k)}\}_{k=1}^K$.
The first constraint encourages the commutativity in \eqref{e:commut} with $\bbP^{(k)}$ as an approximation of $\bbP^{*(k)}=\bbCoh^{(k)}\bbSho^{*(k)}$ to avoid a bilinear formulation.
As will be discussed in Section \ref{S:theory}, the upper bound $\epsilon$ accounts for both the sample covariance submatrix error and the difference between $\bbP^{(k)}$ and $\bbP^{*(k)}$.
Thus, similarly to \cite{rey2022jointb}, we introduce the low-rank matrices $\bbP^{(k)}$ to replace entities that depend on hidden nodes.
However, instead of using the standard convex surrogate for low-rankness given by the nuclear norm, we harness the additional structure on $\bbP^{(k)}$ based on the assumptions in Problem 1 via the $\ell_{2,1}$ norm.

Precisely, the last two terms in the objective apply a group Lasso penalty via the $\ell_{2,1}$ norm~\cite{simon2013sparse}, which evaluates the $\ell_1$ norm of the vector containing the $\ell_2$ norm of each column of the input matrix, that is, $\|\bbP^{(k)}\|_{2,1}=\sum_{i=1}^O \|\bbP^{(k)}_{\cdot,i}\|_2$.
Recall that since $H\ll O$ by \textit{(AS1)}, and given the definition of $\bbP^{*(k)}$,  this matrix is not only low-rank but has sparse columns.
Therefore, applying the $\ell_{2,1}$ norm in the third term in the objective encourages a solution $\bbP^{(k)}$ that is column-sparse.
While low-rank constraints are commonly implemented with the convex nuclear norm penalty~\cite{buciulea2022learning}, where solutions with sparse singular values are sought, we simultaneously promote low-rankness while encouraging column sparsity by the group Lasso penalty.
Additionally, since the networks are assumed to have similar sparsity patterns by \textit{(AS3)}, the support of the matrices $\bbSho^{*(k)}$ will be similar, hence rendering the matrices $\bbP^{*(k)}$ with similar column sparsity patterns.
This is captured by the fourth term in the objective.

As is common with optimization problems for sparse network inference, we introduce a convex relaxation of \eqref{e:l0_prob} that enjoys efficient solvability and theoretical guarantees.
Our convex formulation is
\begin{alignat}{3}&
    \min_{ \{\bbSo^{(k)},\bbP^{(k)}\}_{k=1}^K } ~
    \sum_{k=1}^K \alpha_k \| \bbSo^{(k)} \|_1
     + \sum_{k<k'} \beta_{k,k'} \| \bbSo^{(k)}-\bbSo^{(k')} \|_1
    &\nonumber\\&
    \qquad \qquad \qquad 
     + \sum_{k=1}^K \gamma_k \| {\bbP}^{(k)} \|_{2,1}
     + \sum_{k<k'} \eta_{k,k'} \left\| \left[\begin{matrix}\bbP^{(k)}\\\bbP^{(k')}\end{matrix}\right] \right\|_{2,1}
    &\nonumber\\&
    \st ~\!\!
    \textstyle\sum_{k=1}^K \| \hbCo^{(k)}\bbSo^{(k)} - \bbSo^{(k)}\hbCo^{(k)} + \bbP^{(k)} - (\bbP^{(k)})^\top \|_F^2 \leq \epsilon^2,
    &\nonumber\\&
    \qquad ~\!\!
    \bbSo^{(k)}=(\bbSo^{(k)})^\top, ~ 
    \mathrm{diag}(\bbSo^{(k)}) = {\bf 0}, ~ 
    \forall k=1,\dots, K,
    &\nonumber\\&
    \qquad\!\!
    \textstyle\sum_{j} [\bbSo^{(1)}]_{j1} = 1,
    \label{e:l1_prob}
&\end{alignat}
where we have removed the nonconvexities in \eqref{e:l0_prob} by substituting the $\ell_0$ norms in the objective with convex $\ell_1$ norms.
We further specified the constraints according to \eqref{e:valid_adjmat} for valid adjacency submatrices.
While the last constraint is valid to preclude trivial adjacency submatrices, it would not be viable for graph Laplacians as GSOs.
However, the theoretical results in Section \ref{S:theory} still hold for graph Laplacian GSOs by replacing the last constraint in \eqref{e:valid_adjmat} to enforce valid graph Laplacian submatrices. 

The method presented in \eqref{e:l1_prob} is a convex optimization approach that incorporates the structure from multiple networks and the presence of hidden nodes.
However, it involves estimating $2K$ matrices of size $O \times O$, so its computational complexity is given by $\ccalO(K^{3.5}O^7)$ when an off-the-shelf solver is employed.
While the elevated complexity precludes the application of the proposed method to large graphs, it is possible to design efficient algorithms that exploit the particular problem structure~\cite{percival2012theoretical,yuan2011efficient,obozinski2011group}.
This interesting line of work falls out of the scope of this paper but will be considered as a future line of research.
Finally, recall that the performance of our method is contingent upon a sufficient number of observed nodes compared to the hidden ones, hence assumption \textit{(AS1)}.

\section{Theoretical results}\label{S:theory}

Now, we formalize the viability of the convex relaxation in \eqref{e:l1_prob}.
To that end, we present sufficient conditions under which the solutions to \eqref{e:l0_prob} and \eqref{e:l1_prob} are equivalent even in the presence of hidden nodes.
Moreover, we compute an upper bound on the error of the solution to \eqref{e:l1_prob} that characterizes the effectiveness of the proposed method at accounting for hidden nodes.
Our conclusions follow existing theoretical results for network inference from stationary graph signals~\cite{segarra2017network,navarro2022joint}, but previous works do not consider modifications to the problem such as the inclusion of hidden nodes.
The following results demonstrate that fundamental theoretical guarantees on convex relaxations and error bounds may still be ensured even in the presence of hidden nodes.

\vspace{-.05cm}

\subsection{Sparsity of the convex relaxation}\label{Ss:thm1}
We first introduce the following definitions to rewrite the optimization problems in \eqref{e:l0_prob} and \eqref{e:l1_prob} in vector form.
Let the vectors $\bbalpha\in\mathbb{R}^K$ and $\bbbeta\in\mathbb{R}^{K(K-1)/2}$ collect values of $\alpha_k$ and $\beta_{k,k'}$, respectively.
Let $\ccalL' := \ccalL^{(1)}\cup\cdots\cup\ccalL^{(K)}$, where $\ccalL^{(k)} := \{i = j + (k-1)O^2:j\in\ccalL\}$ for $\ccalL$ containing indices for a $O^2$-length vector (corresponding to the vector form of an $O \times O$ matrix) as described in Section \ref{S:background}.
We define the directed difference matrix $\bbZ := [ {\bf 1}_K^\top \otimes -\bbI_K ]_{\cdot,\ccalL} + [ \bbI_K \otimes {\bf 1}_K^\top ]_{\cdot,\ccalL}$, where $\ccalL$ contains indices for a $K^2$-length vector.
We can then introduce the matrix $\bbPsi := 2[\bbPsi_0]_{\cdot,\ccalL'}$ associated with the objectives of \eqref{e:l0_prob} and \eqref{e:l1_prob}, where
\be
    \bbPsi_0 :=
    \left[\begin{matrix}
        \mathrm{diag}(\bbalpha)\otimes \bbI_{O^2}\\
        \mathrm{diag}(\bbbeta)\bbZ^\top\otimes \bbI_{O^2}
    \end{matrix}\right].
\ee 
For the first constraint of \eqref{e:l0_prob} and \eqref{e:l1_prob}, we introduce $\bbSigma:=\mathrm{blockdiag}(\bbSigma^{(1)},\dots,\bbSigma^{(K)})$, where $\bbSigma^{(k)}:= [\bbSigma_0^{(k)}]_{\cdot,\ccalL} + [\bbSigma_0^{(k)}]_{\cdot,\ccalU}$ and $\bbSigma_0^{(k)} = (-\hbCo^{(k)}\oplus\hbCo^{(k)})$ for all $k=1,\dots,K$, and $\ccalL$ and $\ccalU$ for $\bbSigma^{(k)}$ return entries of a vector of length $O^2$.
Furthermore, let $\bbQ$ be a commutation matrix such that for any square matrix $\bbY$, we have that $\mathrm{vec}(\bbY^\top) = \bbQ\mathrm{vec}(\bbY)$, and let $\bbM = \mathrm{blockdiag}(\bbI_{O^2} - \bbQ,\dots,\bbI_{O^2} - \bbQ)$ with $K$ diagonal blocks.
Let $\ccalE^{(k,i)} = \{ (k-1)O^2 + (i-1)O + j \}_{j=1}^O$ be index sets for all $k=1,\dots,K$ and $i=1,\dots,O$.
Based on this, define $\ccalE^{(k,k',i)}=\ccalE^{(k,i)}\cup \ccalE^{(k',i)}$ for every $k,k'=1,\dots,K$ with $k<k'$, where $\ccalE^{(k,i)}$ corresponds to the indices of the $i$-th column in the vectorized version of the matrix $\bbP^{(k)}$ and $\ccalE^{(k,k',i)}$ to the indices of the $i$-th columns of the vectorized versions of $\bbP^{(k)}$ and $\bbP^{(k')}$.
We combine these sets $\ccalE := \bigcup_{i=1}^O \{\ccalE^{(k,i)}\}_{k=1}^K \cup \{\ccalE^{(k,k',i)}\}_{k<k'}$ and define the set of parameters $\{\eta'_g\}_{g\in\ccalE}$ where $\eta'_{\ccalE^{(k,i)}}=\gamma_k$ and $\eta'_{\ccalE^{(k,k',i)}}=\eta_{k,k'}$ for every $k,k'=1,\dots,K$ such that $k< k'$ and $i=1,\dots, O$.

With the following vectorizations,
\begin{alignat}{3}&
    \bbs = [ \mathrm{vec}(\bbSo^{(1)})_{\ccalL}^\top,~\cdots,~\mathrm{vec}(\bbSo^{(K)})_{\ccalL}^\top ]^\top\in\mathbb{R}^{KO(O-1)/2},
    &\label{e:vecs_s}\\&
    \bbp = [ \mathrm{vec}(\bbP^{(1)})^\top,~\cdots,~\mathrm{vec}(\bbP^{(K)})^\top ]^\top\in\mathbb{R}^{KO^2},
    &\label{e:vecs_p}
\end{alignat}
we may rewrite the optimization problem \eqref{e:l0_prob} as
\begin{alignat}{3}&
    \!\!\!\!
    \{\bbs',\bbp'\} \in \argmin_{\{\bbs,\bbp\}} ~~
    \|\bbPsi\bbs\|_0 + 
    \sum_{g\in\ccalE} \eta_g' \|\bbp_g\|_2
    % \sum_{k=1}^K \sum_{i=1}^O \gamma_k\|\bbp_{\ccalE^{(k,i)}}\|_2
    % &\nonumber\\&
    % \qquad \qquad \qquad \qquad \qquad
    % + \sum_{k<k'} \sum_{i=1}^O \eta_{k,k'} \|\bbp_{\ccalE^{(k,k',i)}}\|_2
    &\nonumber\\&
    \!\!\!\!\!
    \qquad \qquad 
    % \qquad \qquad \quad
    \st
    \|\bbSigma\bbs + \bbM\bbp\|_2 \leq \epsilon,
    ~(\bbe_1\otimes{\bf 1}_{O-1})^\top\bbs = 1
    \tag{\ref{e:l0_prob}'}\label{e:l0_vec_prob}
&\end{alignat}
and \eqref{e:l1_prob} as
\begin{alignat}{3}&
    \!\!\!\!
    \{\hat{\bbs},\hat{\bbp}\} \in \argmin_{\{\bbs,\bbp\}} ~~
    \|\bbPsi\bbs\|_1 +
    \sum_{g\in\ccalE} \eta_g' \|\bbp_g\|_2
    % \sum_{k=1}^K \sum_{i=1}^O \gamma_k\|\bbp_{\ccalE^{(k,i)}}\|_2
    % &\nonumber\\&
    % \qquad \qquad \qquad \qquad \qquad
    % + \sum_{k<k'} \sum_{i=1}^O \eta_{k,k'} \|\bbp_{\ccalE^{(k,k',i)}}\|_2
    &\nonumber\\&
    \!\!\!\!\!
    \qquad \qquad 
    % \qquad \qquad \quad
    \st
    ~~
    \|\bbSigma\bbs + \bbM\bbp\|_2 \leq \epsilon,
    ~(\bbe_1\otimes{\bf 1}_{O-1})^\top\bbs = 1,
    \tag{\ref{e:l1_prob}'}\label{e:l1_vec_prob}
&\end{alignat}
where we do not guarantee unique solutions but consider a given global minimum for~\eqref{e:l0_vec_prob} and~\eqref{e:l1_vec_prob}.
Note that the first term in~\eqref{e:l0_vec_prob} is not equivalent to the first two summations in~\eqref{e:l0_prob} since the $\ell_0$ norm is not homogeneous, but we shall treat them as equivalent for ease of notation.
We further denote $\ccalJ$ as $\mathrm{supp}(\bbPsi\bbs')$ and $\ccalI$ as $\mathrm{supp}(\bbs')$, where $\mathrm{supp}(\bby)$ denotes the support of the vector $\bby$.
With the above definitions in place, we have the following result.

\medskip
\noindent{\bf Theorem 1.} \textit{
Let $\ccalM := \{O,O+1,\dots,KO(O-1)/2\}$.
Assume that problems~\eqref{e:l0_vec_prob} and~\eqref{e:l1_vec_prob} are feasible, and that there exist constants $r',\hat{r} > 0$ such that 
\begin{alignat}{3}&
    \epsilon
    &~\geq~&
    \sigma_\mathrm{\max}(\bbSigma)r' + 2 \hat{r} 
    &\nonumber\\&
    &&
    + \sqrt{2}(\sigma_\mathrm{\max}(\bbSigma) + 2) (\|\bbs'\|_2 + \|\bbp'\|_2 - r),
    \nonumber
&\end{alignat}
for $r \in [0, 2^{-1/2} \min\{ \|\bbs'_{\ccalM}\|_2 + \|\bbp'\|_2, r' \})$. 
If the following two conditions are satisfied:
\bi[\it 1)] $\bbSigma_{\cdot,\ccalI}$ is full column rank; and
\i[\it 2)] There exists a constant $\psi>0$ such that
    \be
        % \!\!\!\!\!\!\!\!\!\!\!\!\!\!\!
        \!\!\!\!\!\!\!
        \|\bbPsi_{{\ccalJ^c},\cdot}
        (\psi^{-2}\bbT + \bbPsi_{{\ccalJ^c},\cdot}^\top\bbPsi_{{\ccalJ^c},\cdot})^{-1}
        \bbPsi_{{\ccalJ},\cdot}^\top\|_{\infty} < 1,
    \ee
    where 
    \begin{alignat}{3}&
        % \!\!\!\!\!\!\!\!\!\!\!\!
        \bbT:= {\bbSigma}^\top {\bbSigma} + (\bbe_1\otimes {\bf 1}_{O-1} ) (\bbe_1\otimes {\bf 1}_{O-1} )^\top,
        &\nonumber
    \end{alignat}
\ei
then there exists a solution $\{\hat{\bbs},\hat{\bbp}\}$ of \eqref{e:l1_vec_prob} that is equivalent to a solution $\{\bbs',\bbp'\}$ of \eqref{e:l0_vec_prob}. 
}
\medskip

The proof of Theorem 1 can be found in Appendix \ref{A:thm1_proof}, but we also provide a summary here.
To decouple the joint optimization of $\bbs$ and $\bbp$, we consider an alternating minimization algorithm, permitting separate analysis of $\bbs$-subproblems and $\bbp$-subproblems at each iteration.
Proximal alternating minimization~\cite{attouch2010proximal}, an iterative optimization algorithm, applied to \eqref{e:l0_vec_prob} and \eqref{e:l1_vec_prob} can be shown to converge to the original solutions $\{\bbs',\bbp'\}$ and $\{\hat{\bbs},\hat{\bbp}\}$, respectively.
We then show that for a sufficiently large $\epsilon\geq 0$, we can ensure that the $\bbs$-subproblems for \eqref{e:l0_vec_prob} and \eqref{e:l1_vec_prob} are equivalent under the conditions of Theorem 1.
When the iterations grow sufficiently large for convergence, this implies that $\bbs'=\hat{\bbs}$.

Under the sufficient conditions of Theorem 1, the convex relaxation in \eqref{e:l1_prob} enjoys recovery of the sparsest solution of \eqref{e:l0_prob} even in the presence of hidden nodes.
Note that this result differs significantly from that of Theorem 1 in \cite{navarro2022joint} due to the presence of another variable $\bbp$ that is not associated with an entrywise sparsity penalty.
Condition \textit{1)} of Theorem 1 guarantees that the solution to \eqref{e:l1_prob} is unique for a fixed $\{\bbP^{(k)}\}_{k=1}^K$, and condition \textit{2)} permits the existence of a dual certificate that ensures that the solutions to \eqref{e:l1_prob} and \eqref{e:l0_prob} are equivalent when $\{\bbP^{(k)}\}_{k=1}^K$ is fixed for both problems~\cite{navarro2022joint,zhang2016one}.
More specifically, if condition \textit{1)} of Theorem 1 holds, then matrices $\bbPsi$ and $\bbSigma$ yield an identifiable optimal solution to~\eqref{e:l1_vec_prob} if $\bbp$ is fixed. 
By the definitions of $\bbPsi$ and $\bbSigma$, this implies that there is only one sparsest set of GSO submatrices that satisfies the relaxed commutativity constraint in~\eqref{e:l0_prob} and~\eqref{e:l1_prob} for a fixed $\{\bbP^{(k)}\}_{k=1}^K$.
Condition \textit{2)} guarantees that the solution $\{\bbs',\hat{\bbp}\}$ is indeed optimal for~\eqref{e:l1_vec_prob}.
Note that when $\bbSigma$ is full column rank, then $\bbT$ is invertible, and we may select an arbitrarily small $\psi>0$ that satisfies condition \textit{2)}~\cite{navarro2022joint}.
In our synthetic simulations, we infer networks from sample covariance submatrices $\hbCo^{(k)}$ obtained from stationary graph signals as graph filter outputs from Gaussian white noise inputs. 
Thus, in all cases of our synthetic simulations, we observe full rank $\hbCo^{(k)}$ and thus full column rank $\bbSigma$, so both conditions \textit{1)} and \textit{2)} hold.
Under the conditions of Theorem 1, the $\ell_1$ norm does not introduce any estimation error for obtaining the sparsest GSO submatrix estimates, and we need only consider the distortion from the sample covariance submatrices $\{\hbCo^{(k)}\}_{k=1}^K$ and auxiliary matrices $\{\hat{\bbP}^{(k)}\}_{k=1}^K$ obtained from \eqref{e:l1_prob}.

\subsection{Robust recovery under hidden nodes}\label{Ss:thm2}

By Theorem 1, we can guarantee under mild conditions when the solution to \eqref{e:l1_prob} is equivalent to the sparsest solution from \eqref{e:l0_prob}.
Therefore, to evaluate the efficacy of our method in estimating the target GSO submatrices $\{\bbSo^{*(k)}\}_{k=1}^K$, we need only consider the estimation error of \eqref{e:l1_prob}.
In the sequel, we derive an upper bound on the distortion between the target GSO submatrices $\{\bbSo^{*(k)}\}_{k=1}^K$ and the estimated ones $\{\hbSo^{(k)}\}_{k=1}^K$ obtained from \eqref{e:l1_prob}.
Let $\bbs^*$ be the vectorization of the target GSO submatrices $\{\bbSo^{*(k)}\}_{k=1}^K$ as in \eqref{e:vecs_s}.
We define $\ccalK$ as $\mathrm{supp}(\bbPsi\bbs^*)$, and we let $R := \sum_{k=1}^K R_k$ and $\omega:=\max_{k=1,\dots,K}\omega_k$, where $\omega_k:=\max\{\max_i[\bbCo^{(k)}]_{ii},\max_i[\bbSo^{*(k)}\bbCo^{(k)}\bbSo^{*(k)}]_{ii}\}$.
We present our main result on the performance of our proposed method.

\medskip

\noindent{\bf Theorem 2.} \textit{
Let $\{\hbSo^{(k)}\}_{k=1}^K$ be the estimated subnetworks obtained from \eqref{e:l1_prob} with $\epsilon = \epsilon_R+\alpha$ for
\be
    \alpha^2 = \sum_{k=1}^K\left\|\big(\hat{\bbP}^{(k)}-(\hat{\bbP}^{(k)})^\top\big)-\big(\bbP^{*(k)}-(\bbP^{*(k)})^\top\big)\right\|_F^2
\ee
and $\epsilon_R \geq C_1 O\omega\sqrt{(K\log O)/R}$ for some constant $C_1 >0$.
Furthermore, assume that observations $\bbX^{(k)}$ correspond to independent realizations of a Gaussian process that is stationary in $\bbS^{(k)}$.
Under the following four conditions,
    \bi[\it 1)] $K = o(\log O)$;
    \i[\it 2)] $R_1 \asymp R_2 \asymp \cdots \asymp R_K $;
    \i[\it 3)] $\log O = o(\min\{ R/(K^7(\log R)^2), (R/K^7)^{1/3} \})$; and
    \i[\it 4)] $\bbSigma$ is full column rank;
    \ei
with probability at least $1-e^{-C_2\log O}$ for some constant $C_2$ we have that 
\begin{alignat}{3}&
    \sum_{k=1}^K \| \hbSo^{(k)} -\bbSo^{*(k)} \|_1 \leq 
    \tau(\epsilon_R + \alpha),
    &\nonumber\\&
    \mathrm{\it where~} 
    \tau = 
    \frac{4\sqrt{|\ccalK|}\sigma_{\mathrm{max}}(\bbPsi)\|\bbPsi^{\dagger}\|_1}{\sigma_{\mathrm{min}}(\bbSigma)}(2+\sqrt{|\ccalK|}).
    \label{e:thm2}
&\end{alignat}
}
\medskip
% \vspace{-.5cm}

The proof of Theorem 2 can be found in Appendix \ref{B:thm2_proof}.
In brief, we first apply the commutative relationship described in Section \ref{S:background} to show that $\{\bbs^*,\hat{\bbp}\}$ is a feasible solution to \eqref{e:l1_vec_prob}.
We can then bound the $\ell_1$-norm difference between the vectorization of the target GSOs  $\bbs^*$ and the estimated one $\hat{\bbs}$ based on the commutativity constraint, $\epsilon=\epsilon_R + \alpha$.

Theorem 2 presents an upper bound on the estimation error of \eqref{e:l1_prob}.
If $K$ and $O$ are fixed, then as the number of observed graph signals $R$ increases, the sample covariance submatrices $\{\hbCo^{(k)}\}_{k=1}^K$ approach the true covariance submatrices, and the first term $\tau\epsilon_R$ in the upper bound in \eqref{e:thm2} becomes negligible.
With enough observed graph signals, the error primarily depends on the second term $\tau\alpha$, which denotes the approximation error of $\{\hat{\bbP}^{(k)}\}_{k=1}^K$, the crux of our proposed method.
The value of $\tau$ depends on the sparsity and similarity of the target GSO submatrices $\bbSo^{*(k)}$ via the presence of $|\ccalK|=|\mathrm{supp}(\bbPsi\bbs^*)|$ and $\sigma_\mathrm{max}(\bbPsi)\|\bbPsi^\dagger\|_1$, which is proportional to the squared condition number of $\bbPsi$.
Additionally, if $\epsilon_R + \alpha$ is less than the lower bound of $\epsilon$ in Theorem 1, then we cannot simultaneously guarantee the results of Theorems 1 and 2. That is, we may adhere to the error bound in Theorem 2 but cannot ensure obtention of the sparsest solution $\{\bbSo^{'(k)}\}_{k=1}^K$.

% \blue{When the true matrices $\{\bbP^{*(k)}\}_{k=1}^K$ have more zero-valued columns with more similar column sparsity patterns, then increasing $\gamma_k$ and $\eta_{k,k'}$ will result in more accurate estimates $\{\hbP^{(k)}\}_{k=1}^K$.
% This will also improve our estimates $\{\hbSo^{(k)}\}_{k=1}^K$ according to Theorem 2. }%
% \blue{However,} w
While characterizing the error of $\hbP^{(k)}$ is of interest, we restrict our analysis to formalizing the solution of the estimate $\hbSo^{(k)}$, which is the problem at hand.
% While characterizing the error of $\hbP^{(k)}$ is of interest, we restrict our analysis to formalizing the solution of the estimate $\hbSo^{(k)}$, which is the problem at hand.
Indeed, quantifying the quality of the estimate $\hbP^{(k)}$ requires knowledge of the error of $\hbSo^{(k)}$, and formalizing the errors of both $\hbP^{(k)}$ and $\hbSo^{(k)}$ is a more challenging and ill-posed setting (both practically and from a theoretical point of view) that goes beyond the scope of this paper.
We instead characterize the error of $\hbSo^{(k)}$ based on the accuracy of the estimate $\hbP^{(k)}$.
If \eqref{e:l1_prob} is effective at enforcing $\bbP^{(k)}$ to share structural characteristics of $\bbCoh^{(k)}\bbSho^{*(k)}$ such that they are close, then the estimation of the GSO submatrices $\bbSo^{*(k)}$ becomes easier according to \eqref{e:thm2}.
Furthermore, as $\bbP^{(k)}$ becomes a more accurate approximation of $\bbP^{*(k)}$, the estimation accuracy of $\hbSo^{(k)}$ improves increasingly when compared to estimating $\bbSo^{*(k)}$ while ignoring the presence of hidden nodes.
We formalize this statement in the following result that characterizes the effectiveness of our proposed formulation with respect to the auxiliary matrices $\{\bbP^{(k)}\}_{k=1}^K$.

\medskip
\noindent{\bf Corollary 1.}\textit{
Let the naive subnetwork estimates considering only observed nodes be denoted as $\{\tbSo^{(k)}\}_{k=1}^K$~\cite{navarro2022joint}, which we define as the solution to~\eqref{e:l1_prob} while fixing $\bbP^{(k)}={\bf 0}_{O\times O}$ for every $k=1,2,\dots,K$, and we let $\epsilon=\epsilon_R$, where $\epsilon_R \geq C_1 O\omega\sqrt{(K\log O)/R}$ for some constant $C_1>0$, and $\gamma_k=0$, $\eta_{k,k'}=0$ for every $k,k'=1,2,\dots,K$ and $k < k'$.
Additionally, let $\tilde{\bbs}$ be the vectorization as in \eqref{e:vecs_s} of $\{\tbSo^{(k)}\}_{k=1}^K$ and define $\delta$ as
\be
\delta^2 = \sum_{k=1}^K \| \bbP^{*(k)} - (\bbP^{*(k)})^\top \|_F^2.
\ee
% Then, we have that
Then, with the same probability and $\tau$ as in Theorem 2
\begin{alignat}{3}&
    \!
    \sum_{k=1}^K \| \tbSo^{(k)} - \bbSo^{*(k)} \|_1 \leq 
    (\tau+\tau')(\epsilon_R + \tfrac{1}{2}\delta),
    &\nonumber\\&
    \!
    \mathrm{\it where~} 
    \tau' = \frac{2\rho KO(O-1) (1+\sqrt{|\ccalK|})\sigma_{\mathrm{max}}(\bbPsi)\|\bbPsi^\dagger\|_1}{\sigma_{\mathrm{min}}(\bbSigma)}
    \label{e:corol_p1}
&\end{alignat}
for some $\rho\in[0,1]$.
Furthermore, we have that if 
\begin{alignat}{3}&
    \sum_{k=1}^K\left\|\big(\hat{\bbP}^{(k)}-(\hat{\bbP}^{(k)})^\top\big)-\big(\bbP^{*(k)}-(\bbP^{*(k)})^\top \big)\right\|_F^2
    &\nonumber\\&
    \quad\quad
    \leq
    \left(\frac{\tau'}{\tau}\right)^2\epsilon_R^2
    + 
    \left(\frac{\tau+\tau'}{2\tau}\right)^2
    \sum_{k=1}^K \left\| \bbP^{*(k)} - (\bbP^{*(k)})^\top \right\|_F^2,
    \label{e:corol_pt2}
&\end{alignat}
then the error bound in \eqref{e:thm2} is lower than that of \eqref{e:corol_p1}.}

\medskip

The proof of Corollary 1 can be found in Appendix \ref{C:coroll_proof}, which follows a similar procedure to the proof of Theorem 2.
Corollary 1 demonstrates the criticality of accounting for hidden nodes.
We describe these implications more intuitively here. 
First, as discussed following Theorem 2, we note that as $\hat{\bbP}^{(k)}$ approximates $\bbP^{*(k)}$ more accurately, we achieve greater improvement over $\{\tbSo^{(k)}\}_{k=1}^K$ from our proposed inference problem \eqref{e:l1_prob}.
Indeed, as the matrix difference $(\hat{\bbP}^{(k)})^\top-\hat{\bbP}^{(k)}$ approaches the right-hand side of \eqref{e:commut}, we remove the influence of the hidden nodes on the estimation of the observed submatrices. 
% \blue{As noted in the discussion following Theorem 2, as $\{\bbP^{*(k)}\}_{k=1}^K$ becomes more column sparse with more similar column support, our estimates $\{\hbP^{(k)}\}_{k=1}^K$ improve with larger $\gamma_k$ and $\eta_{k,k'}$, reducing the left-hand side of~\eqref{e:corol_pt2}.
% }%
Second, note that the second term in the upper bound of \eqref{e:corol_pt2} is proportional to $\delta$, which measures the influence of the hidden nodes on the observed nodes in the stationary graph signal regime.
When $\delta$ is negligible, the hidden nodes have little effect on the observed nodes, and the inclusion of $\{{\bbP}^{(k)}\}_{k=1}^K$ in the inference process may affect performance detrimentally. 
However, as $\delta$ increases, the need to account for the right-hand side of \eqref{e:commut} becomes crucial.
We verify this comparison of \eqref{e:l1_prob} and the naive solution $\{\tbSo^{(k)}\}_{k=1}^K$ with synthetic simulations in Section \ref{S:results}.

Theorem~2 and Corollary~1 demonstrate the error bound for $\hbSo^{(k)}$ in terms of how well we can account for hidden nodes, that is, the estimation error of $\hbP^{(k)}$.
These results demonstrate the need to address missing nodes, as we can achieve error rates similar to a setting with fully observed graph signals if we accurately estimate $\bbP^{*(k)}$.
However, we may wish to characterize the error of $\hbSo^{(k)}$ with respect to the hidden node effect encoded in $\bbP^{*(k)}$ to determine in which problem settings we can achieve accurate estimations.
To this end, we extend the results in Theorem~2 and Corollary~1 to consider the error bound of $\hbSo^{(k)}$ in terms of $\bbP^{*(k)}$, without requiring $\hbP^{(k)}$.

We require additional notation for our next result.
First, recall that by the definitions of the index set $\ccalE$ and the parameter set $\{\eta_g'\}_{g\in\ccalE}$ we have the equivalence
\begin{equation*}
    \sum_{g\in\ccalE}\eta_g' \|\bbp_g^*\|_2
    =
    \sum_{k=1}^K \gamma_k \|\bbP^{*(k)}\|_{2,1}
    +
    \sum_{k<k'} \eta_{k,k'} \left\| \begin{bmatrix}\bbP^{*(k)} \\ \bbP^{*(k')}\end{bmatrix} \right\|_{2,1},
\end{equation*}
and we let $\eta_\mathrm{min}:=\min_{g\in\ccalE} \eta_g'$ and $\eta_\mathrm{max} := \max_{g\in\ccalE} \eta_g'$. 
We also let $\sigma := \sigma_\mathrm{max}(\bbSigma)$ be the largest singular value of the matrix $\bbSigma$.
Our result on the error of $\hbSo^{(k)}$ in terms of the hidden nodes encoded in $\bbP^{*(k)}$ is as follows.

\medskip

\noindent\textbf{Theorem 3.}
\textit{
There exists a constant $\mu>0$ such that when
    \begin{alignat}{3}&
        \!
        \eta_\mathrm{min}
        >
        128\frac{O}{K^2} \mu\tau\sigma \sum_{g\in\ccalE} \|\bbp_g^*\|_2
        + 16 \frac{\sqrt{O}}{K} \mu\epsilon_R (2\tau\sigma+1),
        &\label{e:etamin_lowbnd}\\&
        \qquad\quad
        \eta_\mathrm{max} < 
        \frac{( K\eta_\mathrm{min} - 8\mu\epsilon_R \sqrt{O} (2\tau\sigma + 1) )^2}{ 128O\mu\tau\sigma \sum_{g\in\ccalE} \|\bbp_g^*\|_2 },
        \label{e:etamax_uppbnd}
    &\end{alignat}
then with the same probability and $\tau$ as in Theorem 2 and $\tau'$ as in Corollary 1
    \begin{alignat}{3}&
        \sum_{k=1}^K \| \hbSo^{(k)} -\bbSo^{*(k)} \|_1 \leq 
        \tau(\epsilon_R + \kappa),
        &\nonumber\\&
        \mathrm{\it where~} 
        \kappa = 
        \sqrt{
        \frac{ \sum_{g\in\ccalE} \eta_g' \|\bbp^*_g\|_2 }
             { 2\mu\tau\sigma }}.
        \label{e:thm3}
    &\end{alignat}
Moreover, if
\begin{alignat}{3}&
    \sum_{g\in\ccalE} \|\bbp_g^*\|_2 <
    \frac{K\tau'\epsilon_R}{8\tau\sqrt{O}},
    &\label{e:grouplasso_uppbnd}\\&
    \delta^2 > \frac{4\tau'\epsilon_R^2(2\tau\sigma+1)}{\sigma(\tau+\tau')^2},
    \label{e:delta_lowbnd}
&\end{alignat}
then there exist parameters $\eta_\mathrm{min} \leq \eta_\mathrm{max}$ satisfying both assumptions~\eqref{e:etamin_lowbnd} and~\eqref{e:etamax_uppbnd} and also
\begin{equation}\label{e:etamax_thm3}
    \eta_\mathrm{max}
    <
    \frac{2\mu\sigma}{\tau\sum_{g\in\ccalE}\|\bbp_g^*\|_2}
    ( (\tau'\epsilon_R)^2 + 
    {\textstyle\frac{1}{4}} (\tau+\tau')^2\delta^2),
\end{equation}
which guarantees that the error bound for $\hbSo^{(k)}$ in~\eqref{e:thm3} is lower than that of the naive solution in~\eqref{e:corol_p1}.
If $\alpha\leq\kappa$, then this ensures that~\eqref{e:corol_pt2} holds.
}%

\medskip

Appendix~\ref{D:thm3_proof} presents the proof of Theorem 3.
In short, we first obtain an upper bound for the error of $\hbp$, which we then use to bound the error of $\hbs$ via~\eqref{e:thm2}.
As the true matrices $\bbP^{*(k)}$ increase in column sparsity and similarity, $\kappa$ and thus the error bound decreases.
Moreover, the ratio $\eta_\mathrm{max}/\mu$ tunes between emphasizing the strongest group lasso penalty and the commutativity constraint.
Thus, its presence in $\kappa$ indicates that when $\bbP^{*(k)}$ has denser and more dissimilar column sparsity patterns, we decrease the ratio $\eta_\mathrm{max}/\mu$, reducing encouragement of the group lasso penalties to maintain the error bound. 
Note that $\mu$ is negatively correlated with $\epsilon_R + \alpha$; if $\epsilon_R + \alpha$ is small, then $\mu$ will be large.
However, in this case there may be no feasible ratio $\eta_\mathrm{min}/\mu$ satisfying~\eqref{e:etamin_lowbnd}.
Thus, the right-hand side of~\eqref{e:etamin_lowbnd} must be small enough to ensure a valid lower bound for $\eta_\mathrm{min}/\mu$.
For instance, we may reduce the magnitude of the sparsity parameters $\alpha_k$ and $\beta_{k,k'}$ to reduce $\tau$, which permits more values of $\eta_\mathrm{min}/\mu$ that satisfy~\eqref{e:etamin_lowbnd}.

To guarantee that our method improves the error bound of the naive method in Corollary 1, we require two bounds.
First, we have an upper bound on $\sum_{g\in\ccalE} \|\bbp_g^*\|_2$, that is, the column sparsity patterns of $\{\bbP^{*(k)}\}_{k=1}^K$ must adhere closely enough to our hypothesis.
Second, the effect of $\bbP^{*(k)}$ on~\eqref{e:commut} must be large enough to warrant its estimation via $\hbP^{(k)}$.
When there is no effect due to hidden nodes, that is, $\delta=0$, then it may be unhelpful to mitigate its effect by estimating $\hbP^{(k)}$.
The upper bound~\eqref{e:etamax_thm3} restricts how strongly we enforce the group lasso penalties. 
As $\sum_{g\in\ccalE}\|\bbp_g^*\|_2$ increases or $\delta$ decreases, the largest $\eta_\mathrm{max}$ that guarantees an improved error rate decreases.
Thus, we demonstrate the conditions with respect to hidden node behavior, that is, $\sum_{g\in\ccalE}\|\bbp_g^*\|_2$ and $\delta$, for which we may select parameters $\{\eta_g'\}_{g\in\ccalE}$ that guarantee that our estimates $\hbSo^{(k)}$ via~\eqref{e:l1_prob} have a lower error bound than that of the naive solution $\tbSo^{(k)}$, which ignores hidden nodes.

%%%%%%%%%%%%%%%   FIGURES   %%%%%%%%%%%%%%%
\begin{figure*}[!t]
	\centering
	\begin{subfigure}{0.31\textwidth}
		\centering
		\scalebox{.9}{\input{figs/thm2}}
	\end{subfigure}
	\hspace{.1cm}
	\begin{subfigure}{0.31\textwidth}
		\centering
		\scalebox{.9}{\input{figs/n_graphs}}
	\end{subfigure}
	\hspace{.1cm}
	\begin{subfigure}{0.31\textwidth}
		\centering
        \scalebox{.9}{\input{figs/hidden_nodes}}
	\end{subfigure}
	\caption{
    We test the performance of the proposed network topology inference in different settings.
	(a) Evaluation of the performance of graph inference accounting for hidden nodes via \eqref{e:l1_prob} and graph inference ignoring hidden nodes as described in Corollary 1 as the weights of edges between observed and hidden nodes increase.
	(b) Evaluation of the influence of increasing the number of graphs being estimated.
	(c) Evaluation of the detrimental effects of increasing the number of hidden nodes.
	The experiments consider different graph learning alternatives and the reported results are the average error of 100 independent realizations.}
    \label{fig:synthetic_exps}
\end{figure*}
%%%%%%%%%%%%%%%%%%%%%%%%%%%%%%%%%%%%%%%%%%%%%

\section{Numerical evaluation}\label{S:results}
We introduce several experiments to assess the performance of the proposed network topology inference method.
The experiments employ synthetic and real-world data and compare the quality of the graphs estimated by different algorithms.
For the $k$-th graph, we compute the normalized error between the target $\bbSo^{*(k)}$ and the estimated $\hbSo^{(k)}$ as
\begin{equation}\label{eq:nerr}
    \mathrm{nerr}(\bbSo^{*(k)}, \hbSo^{(k)}) = \frac{\| \bbSo^{*(k)} - \hbSo^{(k)} \|_F^2}{\| \bbSo^{*(k)} \|_F^2},
\end{equation}
and then report the average across the $K$ graphs being estimated, i.e., $\frac{1}{K}\sum_{k=1}^K \mathrm{nerr}(\bbSo^{*(k)}, \hbSo^{(k)})$.
% The results report the normalized error (\textit{nerr}) computed as
% \begin{equation}\label{eq:nerr}
%     \frac{1}{K}\sum_{k=1}^K \frac{\| \bbS^{(k)} - \hbS^{(k)} \|_F^2}{\| \bbS^{(k)} \|_F^2},
% \end{equation}
%where $\bbS^{*(k)}$ and $\hbS^{(k)}$ respectively denote the target and the estimated GSO.
The code for the proposed method and the experiments is available on GitHub\footnote{\url{https://github.com/reysam93/hidden_joint_inference}}.
% The interested reader is referred to the online code for implementation details, as well as for additional complementary simulations that, due to space constraints, are not presented in the paper. 

\subsection{Synthetic experiments}\label{Ss:synth}
We rely on synthetic graphs and signals to assess how different elements impact the performance of the proposed approach. 
Unless specified otherwise, in the following experiments we consider $K=3$ graphs with $N=20$ nodes from which $O=19$ are observed.
The graph $\ccalG^{(1)}$ is sampled from an Erd\H{o}s-Rényi (ER) random graph model with a link probability of $p=0.2$, and the related graphs are created by randomly rewiring a fixed number of edges.  
We ensure that sampled graphs are connected to preclude any isolated nodes.
%To generate stationary graph signals, we diffuse a white input signal across the graph, i.e., $\bbx = \bbH\bbw$, where the coefficients of $\bbH$ are drawn from a uniform distribution and $\bbw \sim \ccalN(\bbzero, \bbI)$.
Stationary graph signals are generated by diffusing a white input signal across the graph, that is, $\bbx = \bbH\bbw$, where the coefficients of $\bbH$ are drawn from a uniform distribution and $\bbw \sim \ccalN(\bbzero, \bbI)$.
Under this model, the covariance of $\bbx$ is a polynomial of $\bbS$, which constitutes a more general setting than, for example, graph signals sampled from a GMRF.
We also replace the first constraint in (5) with a penalty in the objective function, whose weight we can increase for stronger constraint on commutativity.

\vspace{1mm}\noindent
% \textbf{Test case 1.}
\textbf{Varying the effect of hidden nodes.}
We start by illustrating the result in \eqref{e:corol_pt2} that expresses when it is beneficial to incorporate $\bbP^{(k)}$ for hidden nodes.
To this end, we estimate $K=3$ networks from perfectly known covariance submatrices $\bbCo^{(k)}$ so $\epsilon_R=0$ [cf. \eqref{e:corol_pt2}], to assess only the effects of $\bbP^{(k)}$ and the hidden nodes $\ccalH$, characterized respectively by $\alpha$ from Theorem 2 and $\delta$ from Corollary 1.
We compare two network inference methods: (i) ``JH-GSR'', which denotes the method in \eqref{e:l1_prob} that accounts for hidden nodes, and (ii) ``J-GSR'', which denotes the method described in Corollary 1 that ignores hidden variables~\cite{navarro2022joint}.
\cref{fig:synthetic_exps}a shows the network estimation error as the edge weights connecting observed nodes and hidden nodes increase, that is, as nonzero entries in $\bbSoh^{*(k)}$ grow larger.
While the GSO sparsity patterns do not change, the hidden node influence $\delta$ increases with the edge weights in $\bbSoh^{*(k)}$.
To measure performance that is consistent with Corollary 1, we report the average error across all $K$ graphs as the normalized $\ell_1$-norm difference, equivalent to computing \eqref{eq:nerr} with the $\ell_1$ norm replacing the squared Frobenius norm.
% We use the same value of $\epsilon=10^{-8}$ for the first constraint in \eqref{e:l1_prob}; however, the naive problem \blue{with $\bbP^{(k)}={\bf 0}_{O\times O}$} may not be feasible.
We let $\epsilon=10^{-8}$ for the first constraint in \eqref{e:l1_prob}; however, the solution to the naive problem with $\bbP^{(k)}={\bf 0}_{O\times O}$ may not be feasible.
Indeed, when $\epsilon$ is small enough, it may be impossible to obtain a feasible solution $\{\tbSo^{(k)}\}_{k=1}^K$ such that all constraints hold.
In such a case where the solution is infeasible, we let its error be 1.
Along with network estimation error, we compare in \cref{fig:synthetic_exps}a normalized values of $\alpha$ and $\delta$ to evaluate when the result in \eqref{e:corol_pt2} holds.
In particular, we let $\bar{\alpha}:=\sum_k \mathrm{nerr}( \bbP^{*(k)}, (\bbP^{*(k)})^\top+\hat{\bbP}^{(k)}-(\hat{\bbP}^{(k)})^\top )/K $ and $\bar{\delta}:=\sum_k \mathrm{nerr}(\bbP^{*(k)},(\bbP^{*(k)})^\top)/K $.
Since we need only consider which value is greater, we plot $\bar{\alpha}/C$ and $\bar{\delta}/C$ for a constant $C>0$ such that the values are between 0 and 1.

When the edge weight is 0, the hidden nodes are decoupled from the network and thus have no effect on the observed nodes, and indeed ``J-GSR'' perfectly recovers the target networks.
For zero-valued edge weights in $\bbSoh^{*(k)}$, we observe $\alpha \geq \delta$, where ``JH-GSR'' is comparable but not superior to ``J-GSR''.
As the edge weight increases and becomes nonnegligible, the effect of the hidden nodes increases, and we observe in \cref{fig:synthetic_exps}a that $\alpha < \delta$ for all nonzero edge weights and ``JH-GSR'' consistently outperforms ``J-GSR'' as expected from \eqref{e:corol_pt2}.
We thus validate the necessity of our proposed method, where as the influence of hidden nodes increases, we must account for their presence to maintain a satisfactory estimation error.

%%%%%%%%%%%%%%%   FIGURES   %%%%%%%%%%%%%%%
\begin{figure*}[!t]
	\centering
	\begin{subfigure}{0.31\textwidth}
		\centering
		\scalebox{.9}{\input{figs/graph_sim}}
	\end{subfigure}
	\hspace{.1cm}
	\begin{subfigure}{0.31\textwidth}
		\centering
		\scalebox{.9}{\input{figs/sparsity}}
	\end{subfigure}
	\caption{
	We test the performance of the proposed method in different scenarios.
	(a) Evaluation of the impact of the graph similarity in joint network topology inference methods in different graph learning alternatives.
	(b) Evaluation of the impact of the graph sparsity in the support recovery for different hyperparameter selections. The experiment considers two settings for graph similarity by rewiring 3 and 6 links. The results reported are the average error of 100 independent realizations.
	} \label{fig:synthetic_aug_exps} 
\end{figure*}
%%%%%%%%%%%%%%%%%%%%%%%%%%%%%%%%%%%%%%%%%%%%%

\vspace{1mm}\noindent
% \textbf{Test case 2.}
\textbf{Varying the number of graphs.}
We next assess the benefits of considering a joint network topology inference approach when several graphs need to be learned.
To that end, \cref{fig:synthetic_exps}b illustrates the normalized error computed according to \eqref{eq:nerr} as the number of graphs $K$ being estimated increases.
The performance of ``JH-GSR'' is compared with (i) ``S-GSR'', the network topology inference method from stationary observations~\cite{segarra2017network} where graphs are learned individually and the presence of hidden variables is ignored; ``SH-GSR'', a generalization of (i) that takes into account the influence of hidden variables~\cite{buciulea2022learning}; and (iii) ``J-GSR'' as in \cref{fig:synthetic_exps}a.
Looking at the results, we observe that ``JH-GSR'' outperforms the alternatives, showcasing the benefits of harnessing the graph similarity while accounting for the influence of the hidden nodes.
We also observed that the joint approaches achieve a lower error when more than one graph is being estimated, and furthermore, that the benefits of the joint approaches increase with $K$. 
Lastly, \cref{fig:synthetic_exps}b also shows that for the setup at hand, ignoring the influence of hidden nodes results in a worse performance than ignoring the relation across networks, which is studied in more detail in the following experiment.

\vspace{1mm}\noindent
% \textbf{Test case 3.}
\textbf{Varying the number of hidden nodes.}
The results in \cref{fig:synthetic_exps}c investigate the detrimental influence of the presence of hidden nodes in the network topology inference task.
We examine fixed-size graphs with $N=20$ nodes and increase the number of hidden nodes $H$ as shown in the x-axis.
We evaluate the performance of (i) our proposed method, ``JH-GSR'', (ii) an alternative implementation of our method replacing the group Lasso penalty by the nuclear norm, ``NN'', and (iii) the joint network topology inference ignoring the presence of hidden nodes, ``J-GSR''~\cite{navarro2022joint}.
Then, for each baseline, we consider the estimation of either 2 or 6 graphs.
First, from \cref{fig:synthetic_exps}c, it can be seen that increasing the number of hidden nodes renders the inference problem more challenging and, moreover, that ignoring the presence of hidden nodes results in poor performance. 
Second, the superior performance of ``JH-GSR'' over ``NN'' supports our initial intuition that the group Lasso penalty is better suited to capture the structure of the problem at hand. 
Furthermore, we also observe that estimating 6 graphs leads to a better performance than estimating 2, a behavior aligned with the previous experiment.  

\vspace{1mm}\noindent
% \textbf{Test case 4.}
\textbf{Varying graph similarity.}
Next, we evaluate the impact of \textit{(AS3)}, a critical assumption in joint graph learning.
More precisely, we consider estimating $K=3$ graphs as the proportion of different edges increases, i.e., as the graphs become more dissimilar. 
The errors of the estimated graphs are depicted in \cref{fig:synthetic_aug_exps}a, where we compare the performance of ``JH-GSR'' with (i) ``LVGL'', a graphical Lasso algorithm modeling the presence of hidden nodes~\cite{chandrasekaran2012latent}; and (ii) ``FGL'', a joint graphical Lasso algorithm~\cite{danaher2014joint}.
Moreover, since graphical Lasso algorithms assume that the observations are drawn from a GMRF, we consider two different types of signals.
Signals sampled from a GMRF are denoted as ``M'', and signals generated as the diffusion of a white input via a polynomial of the GSO are denoted as ``P''.
As expected from \textit{(AS3)}, \cref{fig:synthetic_aug_exps}a shows that the performance of joint methods, ``JH-GSR'' and ``FGL'', deteriorates as we consider a higher number of different links.
For the two signal models, we observe that ``JH-GSR-M'' is superior to ``JH-GSR-P'' since the GMRF model is a simpler special case of graph stationarity that is less sensitive to hidden nodes.
Interestingly, ``JH-GSR-M'' also outperforms ``FGL-M'', although the latter is a method tailored for GMRF observations, showcasing the more general nature of the stationary model and the importance of accounting for the presence of hidden nodes.
In contrast, we observe that graphical models are incapable of estimating graphs from stationary observations, and we note that ``LVGL-P'' is not included in the figure due to its high error.

\vspace{1mm}\noindent
\textbf{Varying graph sparsity.}
In the last experiment based on synthetic data, we assess the performance of the proposed method in terms of the recovery of the support and how the weight of the regularizers influences the results.
To that end, \cref{fig:synthetic_aug_exps}b depicts the evolution of the \textit{F1-score} as the mean node degree increases for different configurations of the hyperparameters.
Graph $\ccalG^{(1)}$ is drawn from a small world random graph model with a rewiring probability of 0.1, and similar graphs are generated by rewiring either 3 or 6 links (respectively ``3Rw'' or ``6Rw'' in the legend).
The results illustrate how higher values of $\alpha$ obtain the best performance when the graph is sparse but deteriorates as the graph becomes denser.
Similarly, a high value of $\beta$ harnesses the similar support of the graphs but, when graphs are less alike, it may deteriorate the performance.
Last but not least, \cref{fig:synthetic_aug_exps}b illustrates how the support of the graphs is almost perfectly recovered when graphs are sparse, but the performance deteriorates as the density of edges increases.
% two different scenarios for graph similarity, with 3 and 6 rewiring links representing more and less similar graphs respectively. The parameter configurations i) [$\alpha=5$, $\beta=100$, $\gamma=300$] and ii) [$\alpha=10$, $\beta=50$, $\gamma=200$] represent the best parameter combinations to maximize the results of support recovery for dense and sparse graphs.
% As can be seen from the results, higher values of $\alpha$ promote sparser graphs and obtain a high estimate of the graph when it is sparse but reducing quickly the quality of the support estimation as the graph is less sparse.
%When we have sparse graphs, we see that a large value of alpha helps the graph to recover correctly and that as the graphs become denser we observe that the fscore is penalised more (by the slope of the red vs. blue lines) when we promote more sparsity. 
%If we focus on the similarity between graphs, a large beta value promotes that the estimated graphs are more similar, that is why the solid lines are above the dotted lines when the graphs are sparse. When having denser graphs this behaviour is altered as the Fscore is penalised by alpha. 
%Finally, a high gamma value reduces/ignores the influence of hidden variables. That is why we can see that the red lines have a better fscore than the blue ones, because gamma is lower and takes into account the presence of hidden variables.   

%%%%%%%%%%%%%%%   FIGURES   %%%%%%%%%%%%%%%
\begin{figure*}[!t]
	\centering
	\begin{subfigure}{0.31\textwidth}
		\centering
		\scalebox{.9}{\input{figs/samples_real}}
	\end{subfigure}
	\hspace{.1cm}
	\begin{subfigure}{0.31\textwidth}
		\centering
		\scalebox{.9}{\input{figs/votes}}
	\end{subfigure}
	\hspace{.1cm}
    \begin{subfigure}{0.31\textwidth}
		\centering
		\scalebox{.9}{\input{figs/votes_4h}}
	\end{subfigure}
	\caption{
	We test the performance of the proposed network topology inference in real-world scenarios.
	(a) Error estimating three graphs considering either a joint or a separate method. Graphs are obtained from the students of the University of Ljubljana dataset.
	(b) Error estimating two graphs from voting signals considering different approaches for $H=2$.
	(c) Error estimating two graphs from voting signals considering different approaches for $H=4$. 
} \label{fig:thm_real_exps} 
\vspace{-.2cm}
\end{figure*}
%%%%%%%%%%%%%%%%%%%%%%%%%%%%%%%%%%%%%%%%%%%%%

\subsection{Application to real-world graphs}\label{Ss:real}
In addition to the synthetic data where we know the model relating the networks and the observed graph signals, we assess our proposed method with real-world data to demonstrate its efficacy in several scenarios, including those where the stationarity assumption is not explicitly enforced.

\vspace{1mm}\noindent
\textbf{Students dataset.}
The following experiment combines real-world graphs with synthetic signals.
This mixed approach allows us to investigate the applicability of the proposed method to real-world graphs while ensuring that the observed signals are stationary.
We employed three graphs defined on a common set of 32 nodes, where nodes represent students from the University of Ljubljana, and the different graphs encode various types of interactions among the students\footnote{{\scriptsize Original data available at  \url{http://vladowiki.fmf.uni-lj.si/doku.php?id=pajek:data:pajek:students}}}.
The results are displayed in  \cref{fig:thm_real_exps}a, where we observe the error of the recovered graphs as the number of samples increases.
The error reported is the average of 50 realizations of random stationary graph signals, with only one hidden node considered.
For each of the three graphs, we evaluate the performance of both the joint and the separate estimation methods, ``JH-GSR'' and ``SH-GSR''.
From the results, it is evident that the recovery of all three graphs significantly improves with a joint approach, demonstrating the benefits of leveraging the existing relationship between the networks.
%Moreover, the experiment provides evidence that our proposed method is also effective in practical scenarios with real-world applications.

\vspace{1mm}\noindent
\textbf{Learning multiple observed graphs from voting data.}
Finally, we close with an experiment aimed at learning two related political graphs from voting data\footnote{\scriptsize Original data available at \url{https://swissvotes.ch/page/home}}.
More specifically, we consider 25 cantons of Switzerland as the nodes of the graph and the percentage of votes in favor of 185 initiatives submitted between 2000 and 2020 as the signals.
In this setting, links reflect social influence between cantons (for example, if a canton has a great influence over others its degree will be larger), and hidden nodes correspond to cantons whose votes are never observed.
Our goal then is to infer the political graph of Switzerland for two consecutive periods of time.
Intuitively, although political representation may evolve with time, this process is typically slow and, hence, the two graphs are expected to be closely related.
We validate the estimations via ground truth graphs whose links reflect the political preferences of the cantons, which are obtained by performing separate inference of both graphs with all available signals.
We consider two setups with $H=2$ and $H=4$ hidden nodes, respectively illustrated in \cref{fig:thm_real_exps}b and \cref{fig:thm_real_exps}c.
The figures present the normalized error of the estimated graphs as the percentage of available signals ranges from 70\% to 90\% of all available signals.
We compare the proposed algorithm, ``JH-GSR'', with three alternative methods: ``J-GSR'', ``SH-GSR'', and ``J-LVGL'' from \cite{rey2022jointb}.

First, we focus on the estimation performance of the four methods when $H=2$ hidden nodes are considered as shown in \cref{fig:thm_real_exps}b.
Since the number of available signals for the second graph is considerably smaller than the signals available for the first graph, we observe a much larger estimation error for the second graph when the separate approach ``SH-GSR'' is employed.
In contrast, for the joint estimation method ``J-GSR'', we observe that errors are similar for both graphs and inferior on average compared to ``SH-GSR''.
This behavior illustrates that harnessing the similarity of the graphs results in an improvement in performance since it allows sharing common learned structures across graphs.
Moreover, we observe that ``JH-GSR'' outperforms both ``SH-GSR'' and ``J-GSR'' since, in addition to being a joint approach, it takes into account the influence of the hidden nodes.
We also compare ``JH-GSR'' with ``J-LVGL'', both of which perform joint network inference while accounting for hidden nodes.
However, we find that ``JH-GSR'' is drastically superior due to complexities in the data structure that ``J-LVGL'' cannot capture accurately.
Indeed, the stationary model subsumes the GMRF model while allowing for more complex statistical relationships between the graph topology and the signals.

Moving to the results of \cref{fig:thm_real_exps}c, we observe that increasing the number of hidden variables renders the problem more challenging, hence leading to a drop in the performance of all the algorithms.
It is worth mentioning that the error corresponding to ``G2 J-LVGL'' was too high, so it is not included in the figure.
Also note that the fraction of hidden nodes is $4/25$, which is relatively large.
Nevertheless, we observe that methods accounting for the presence of hidden nodes are more resilient to this challenging setting, while the performance of the non-robust alternatives deteriorates significantly.
Moreover, the proposed method ``JH-GSR'' continues to outperform the alternatives, achieving a lower error in both recovered graphs.

To summarize, it is not only crucial to account for the presence of hidden nodes but, when several related graphs are involved, it is also important to exploit the similarity between both observed and hidden nodes.
This becomes particularly relevant when data is limited to a subset of the graphs, as demonstrated in the improved estimation of the second graph when considering joint network inference methods.

\section{Conclusion}\label{S:conclusion}

In this paper, we presented a method to infer multiple networks on the same node set in the presence of hidden nodes.
To characterize the effect of the hidden nodes, we assumed that graph signals were stationary on their respective networks.
By the inherent block structure of the covariance matrix $\bbC^{(k)}$ and the GSO $\bbS^{*(k)}$ of the $k$-th network, we introduced a set of auxiliary matrices $\bbP^{(k)}$ to account for the effect of hidden nodes in the relationship $\bbC^{(k)}\bbS^{*(k)}=\bbS^{*(k)}\bbC^{(k)}$ stemming from the stationarity assumption.
By prior assumptions on structure and stationarity, we derive characteristics of $\bbP^{(k)}$ that permit us to form an optimization problem that performs network inference while accounting for the presence of hidden nodes.
Moreover, we verified that the estimation of the sparsest networks is equivalent to a computationally feasible convex relaxation under mild conditions.
We further demonstrated a bound on the error of our proposed method dependent on the error due to the sample covariance matrices and $\bbP^{(k)}$.
The performance of our method was evaluated in multiple synthetic and real-world datasets in comparison with other baseline methods, and we also verified the improvement in estimation due to the incorporation of $\bbP^{(k)}$.

{\appendices
\section{Proof of Theorem 1}\label{A:thm1_proof}

% We first combine the last two terms in the objective functions of \eqref{e:l0_vec_prob} and \eqref{e:l1_vec_prob} by defining the combined index set $\ccalE := \bigcup_{i=1}^O \{\ccalE^{(k,i)}\}_{k=1}^K \cup \{\ccalE^{(k,k',i)}\}_{k<k'}$ and parameters $\{\eta'_g\}_{g\in\ccalE}$ such that $\eta'_{\ccalE^{(k,i)}}=\gamma_k$ and $\eta'_{\ccalE^{(k,k',i)}}=\eta_{k,k'}$ for every $k,k'=1,\dots,K$ such that $k< k'$ and $i=1,\dots, O$.

Let us consider solving \eqref{e:l0_vec_prob} by proximal alternating minimization \cite{attouch2010proximal} with
\begin{subequations}\label{e:l0_am}
\begin{alignat}{3}&
    {\bbp'}^{(t)} = \argmin_{\bbp} \sum_{g\in\ccalE} \eta'_g\|\bbp_g\|_2
    + \frac{1}{2\lambda'_t}\| \bbp - {\bbp'}^{(t-1)} \|_2^2
    &\nonumber\\&
    \qquad \qquad ~
    \st ~
    \|\bbSigma{\bbs'}^{(t-1)} + \bbM\bbp\|_2 \leq \epsilon,
    &\label{e:l0_am_p}\\&
    {\bbs'}^{(t)} \in \argmin_{\bbs} \|\bbPsi\bbs\|_0 
    + \frac{1}{2\mu'_t}\| \bbs - {\bbs'}^{(t-1)} \|_2^2
    &\nonumber\\&
    \qquad \qquad
    \st \|\bbSigma\bbs + \bbM{\bbp'}^{(t)}\|_2 \leq \epsilon,
    ~(\bbe_1\otimes{\bf 1}_{O-1})^\top\bbs = 1,
    \label{e:l0_am_s}
&\end{alignat}
\end{subequations}
and \eqref{e:l1_vec_prob} with
\begin{subequations}\label{e:l1_am}
\begin{alignat}{3}&
    \hat{\bbp}^{(t)} = \argmin_{\bbp} \sum_{g\in\ccalE} \eta'_g\|\bbp_g\|_2
    + \frac{1}{2\hat{\lambda}_t}\| \bbp - \hat{\bbp}^{(t-1)} \|_2^2
    &\nonumber\\&
    \qquad \qquad ~
    \st ~
    \|\bbSigma\hat{\bbs}^{(t-1)} + \bbM\bbp\|_2 \leq \epsilon,
    &\label{e:l1_am_p}\\&
    \hat{\bbs}^{(t)} = \argmin_{\bbs} \|\bbPsi\bbs\|_1
    + \frac{1}{2\hat{\mu}_t}\| \bbs - \hat{\bbs}^{(t-1)} \|_2^2
    &\nonumber\\&
    \qquad \qquad
    \st \|\bbSigma\bbs + \bbM\hat{\bbp}^{(t)}\|_2 \leq \epsilon,
    ~(\bbe_1\otimes{\bf 1}_{O-1})^\top\bbs = 1,
    \label{e:l1_am_s}
&\end{alignat}
\end{subequations}
for $t\in\mathbb{N}$, where the parameters $\lambda'_t$, $\mu'_t$, $\hat{\lambda}_t$, and $\hat{\mu}_t$ are bounded above and below by positive real numbers.
By the proximal terms in~\eqref{e:l0_am} and~\eqref{e:l1_am}, the subproblems~\eqref{e:l0_am_p},~\eqref{e:l1_am_p}, and~\eqref{e:l1_am_s} are strongly convex, and each iteration of these has a unique solution.
Furthermore, for every $t\in\mathbb{N}$ and any given pair of constants $C^s_t,C^p_t\geq 0$, we may select positive values $\lambda'_t$, $\mu'_t$, $\hat{\lambda}_t$, and $\hat{\mu}_t$ such that the solutions to \eqref{e:l0_am} and \eqref{e:l1_am} are equivalent to
% \madcomment{Actually, it's possible for $C^s_t$ or $C^p_t$ to be 0. A ridge penalty with finite weight can achieve full sparsity. For example, a finite nonzero $\lambda'_t$ could result in ${\bbp'}^{(t)}={\bbp'}^{(t-1)}$.}
\begin{subequations}\label{e:l0_am_equiv}
\begin{alignat}{3}&
    {\bbp'}^{(t)} = \argmin_{\bbp} \sum_{g\in\ccalE} \eta'_g\|\bbp_{g}\|_2
    &\nonumber\\&
    \qquad \quad ~
    \st ~
    \|\bbSigma{\bbs'}^{(t-1)} + \bbM\bbp\|_2 \leq \epsilon, ~
    \| \bbp - {\bbp'}^{(t-1)} \|_2 \leq C^p_t,
    &\label{e:l0_am_p_equiv}\\&
    {\bbs'}^{(t)} \in \argmin_{\bbs} \|\bbPsi\bbs\|_0 
    &\nonumber\\&
    \qquad \quad ~ 
    \st ~ \|\bbSigma\bbs + \bbM{\bbp'}^{(t)}\|_2 \leq \epsilon,
    ~(\bbe_1\otimes{\bf 1}_{O-1})^\top\bbs = 1
    &\nonumber\\&
    \qquad \qquad \quad ~~ \!
    \| \bbs - {\bbs'}^{(t-1)} \|_2 \leq C^s_t,
    \label{e:l0_am_s_equiv}
&\end{alignat}
\end{subequations}
and 
\begin{subequations}\label{e:l1_am_equiv}
\begin{alignat}{3}&
    \hat{\bbp}^{(t)} = \argmin_{\bbp} \sum_{g\in\ccalE} \eta'_g\|\bbp_{g}\|_2
    &\nonumber\\&
    \qquad \quad ~
    \st ~
    \|\bbSigma\hat{\bbs}^{(t-1)} + \bbM\bbp\|_2 \leq \epsilon, ~
    \| \bbp - \hat{\bbp}^{(t-1)} \|_2 \leq C^p_t,
    &\label{e:l1_am_p_equiv}\\&
    \hat{\bbs}^{(t)} = \argmin_{\bbs} \|\bbPsi\bbs\|_1
    &\nonumber\\&
    \qquad \quad ~ 
    \st ~ \|\bbSigma\bbs + \bbM\hat{\bbp}^{(t)}\|_2 \leq \epsilon,
    ~(\bbe_1\otimes{\bf 1}_{O-1})^\top\bbs = 1
    &\nonumber\\&
    \qquad \qquad \quad ~~ \!
    \| \bbs - \hat{\bbs}^{(t-1)} \|_2 \leq C^s_t.
    \label{e:l1_am_s_equiv}
&\end{alignat}
\end{subequations}

Let us initialize the proximal alternating minimization steps for \eqref{e:l0_am_equiv} and \eqref{e:l1_am_equiv} with $\bbp_0 := {\bbp'}^{(0)} = \hat{\bbp}^{(0)}$ and $\bbs_0 := {\bbs'}^{(0)} = \hat{\bbs}^{(0)}$.
Note that the objective functions of \eqref{e:l0_vec_prob} and \eqref{e:l1_vec_prob} are semi-algebraic functions \cite{friedland2014some} and thus have the Kurdyka-{\L}ojasiewicz property \cite{attouch2010proximal}.
By \cite[Theorem 3.3]{attouch2010proximal}, there exist constants $r', s'>0$ such that when we let $\|\bbp'-\bbp_0\|_2 + \|\bbs'-\bbs_0\|_2 < r'$ and 
\begin{alignat}{3}&
    % \min_{\bbs,\bbp}
    \|\bbPsi\bbs'\|_0 + \sum_{g\in\ccalE} \eta'_g \|\bbp'_{g}\|_2
    &~\leq~&
    \|\bbPsi\bbs_0\|_0 + \sum_{g\in\ccalE} \eta'_g \|[\bbp_0]_{g}\|_2
    &\nonumber\\&
    &~<~&
    % \min_{\bbs,\bbp}
    \|\bbPsi\bbs'\|_0 + \sum_{g\in\ccalE} \eta'_g \|\bbp'_{g}\|_2 + s',
    &\label{e:pam_s}
&\end{alignat}
where the first inequality is due to the optimality of $\{\bbs',\bbp'\}$ for feasible $\{\bbs_0,\bbp_0\}$, then we have that the sequence $\{{\bbs'}^{(t)},{\bbp'}^{(t)}\}$ converges to $\{\bbs',\bbp'\}$ in finitely many steps. Similarly, there exist constants $\hat{r},\hat{s}>0$ such that we can guarantee that the sequence $\{\hat{\bbs}^{(t)},\hat{\bbp}^{(t)}\}$ converges to $\{\hat{\bbs},\hat{\bbp}\}$ in finitely many steps.
More specifically, there exist positive integers $T_1,T_2$ such that $\{\bbs',\bbp'\}=\{{\bbs'}^{(t)},{\bbp'}^{(t)}\}$ for every $t\geq T_1$ and $\{\hat{\bbs},\hat{\bbp}\}=\{\hat{\bbs}^{(t)},\hat{\bbp}^{(t)}\}$ for every $t\geq T_2$. 
% Under the guaranteed convergence of proximal alternating minimization, we may consider the separated subproblems in \eqref{e:l0_am_equiv} and \eqref{e:l1_am_equiv} to show equivalence of $\{\bbs',\bbp'\}$ and $\{\hat{\bbs},\hat{\bbp}\}$.

Note that $\hat{r}>0$ may take any arbitrarily large finite number~\cite{attouch2010proximal}, so we may select $\{\bbs_0,\bbp_0\}$ such that $\|\bbp'-\bbp_0\|_2 + \|\bbs'-\bbs_0\|_2 < r'$ and~\eqref{e:pam_s} are satisfied.
Then, we let $\hat{r}\geq r' + \|\hat{\bbs}-\bbs'\|_2$.
Such a finite $\hat{r}$ exists since problems~\eqref{e:l0_vec_prob} and~\eqref{e:l1_vec_prob} have coercive objective functions and we assume feasibility of both, that is, $\|\hat{\bbs}-\bbs'\|_2 \leq \|\hat{\bbs}\|_2 + \|\bbs'\|_2 < +\infty$.
Similarly, we may select a finite upper bound $C_s = C^s_t \geq \|\hat{\bbs} - \bbs'\|_2$ for every $t\geq T$ for the last constraint in subproblems~\eqref{e:l0_am_s_equiv} and~\eqref{e:l1_am_s_equiv}.

We select feasible initial points $\{\bbs_0,\bbp_0\}$ to guarantee convergence of~\eqref{e:l0_am} and~\eqref{e:l1_am}.
Recall that we define the set $\ccalM = \{O,O+1,\dots,KO(O-1)/2\}$, and let $\bba' := [{\bbs'_{\ccalM}}^\top,{\bbp'}^\top]^\top$ and $\bba_0 := [[\bbs_0]^\top_{\ccalM},\bbp_0^\top]^\top$.
% \red{[SR: Since we use $t$ for time index, we may want to use a different symbol]}
Consider the optimization problem
\be
    \min_{\bba_0} \|\bba_0\|_2^2 
    ~\st~
    \|\bba'-\bba_0\|_2 \leq r,
    % ~[\bba_0]_i = \bbs'_i~\forall i=1,2,\dots,O-1,
\ee
whose optimal solution is $\bba_0 = C\bba'$ where $C = (\|\bba'\|_2 - r)/\|\bba'\|_2$.
Then, our optimal initial point is $[\bbs_0]_{\ccalM^c} = \bbs'_{\ccalM^c}$, $[\bbs_0]_{\ccalM} = C\bbs'_{\ccalM}$, and $\bbp_0 = C\bbp'$.
By the inequality $(a+b)^2 \leq 2a^2 + 2b^2$ and our assumption that $r < 2^{-1/2}(\|\bbs'_{\ccalM}\|_2 + \|\bbp'\|_2) \leq \|\bba'\|_2$, we have that $C\in[0,1)$.
Moreover, the solution $\{\bbs_0,\bbp_0\}$ satisfies $\|\bbs'-\bbs_0\|_2 + \|\bbp'-\bbp_0\|_2 \leq \sqrt{2}\|\bba'-\bba_0\|_2 \leq \sqrt{2}r < r'$.
By our condition on $\epsilon$, we have that
\begin{alignat}{3}&
    \epsilon
    &~\geq~&
    \sigma_{\mathrm{max}}(\bbSigma) r' + 2 \hat{r}
    &\nonumber\\& &&
    + \sqrt{2}( \sigma_{\mathrm{max}}(\bbSigma) + 2 )
    (\|\bbs'\|_2 + \|\bbp'\|_2 - r)
    &\nonumber\\&
    &~\geq~&
    \sigma_{\mathrm{max}}(\bbSigma) r' + 2 \hat{r}
    &\nonumber\\& &&
    + ( \sigma_{\mathrm{max}}(\bbSigma) + 2 )
    ( \|\bbs'_{\ccalM^c}\|_2 + C\sqrt{2}\|\bba'\|_2)
    &\nonumber\\&
    &~\geq~&
    \sigma_{\mathrm{max}}(\bbSigma) r' + 2 \hat{r}
    &\nonumber\\& &&
    + ( \sigma_{\mathrm{max}}(\bbSigma) + 2 )
    (\|\bbs_0\|_2 + \|\bbp_0\|_2).
    \nonumber
&\end{alignat}
Then, since $\sigma_{\mathrm{max}}(\bbM) = 2$, 
\begin{alignat}{3}&
    \|\bbSigma\bbs' + \bbM\hat{\bbp}\|_2
    &~\leq~&
    \|\bbSigma(\bbs'-\bbs_0)\|_2
    +
    \|\bbM(\hat{\bbp}-\bbp_0)\|_2
    &\nonumber\\&
    && \quad
    +
    \|\bbSigma\bbs_0 + \bbM\bbp_0\|_2
    &\nonumber\\&
    &~\leq~&
    \sigma_\mathrm{max}(\bbSigma)r' + 2\hat{r} 
    &\nonumber\\&
    && \quad
    + (\sigma_\mathrm{max}(\bbSigma)+2)
    (\|\bbs_0\|_2 + \|\bbp_0\|_2)
    &\nonumber\\&
    &~\leq~&
    \epsilon.
    \label{e:l0_l1_model}
&\end{alignat}

By the finite convergence of~\eqref{e:l0_am_equiv} and~\eqref{e:l1_am_equiv}, we have that $\bbs'={\bbs'}^{(t)}$ and $\hat{\bbs} = \hat{\bbs}^{(t)}$ for every $t\geq T$.
We may rewrite~\eqref{e:l0_am_s_equiv} and~\eqref{e:l1_am_s_equiv} at iteration $T+1$ as
\begin{alignat}{3}&
    {\bbs'} = \argmin_{\bbs} \|\bbPsi\bbs\|_0 
    &\nonumber\\&
    \qquad
    \st \|\bbSigma\bbs + \bbM{\bbp'}\|_2 \leq \epsilon,
    ~(\bbe_1\otimes{\bf 1}_{O-1})^\top\bbs = 1,
    &\nonumber\\&
    \qquad \qquad \!
    \| \bbs - \bbs' \|_2 \leq C_s,
    &\label{e:l0_am_s_laststep}\\&
    \hat{\bbs} = \argmin_{\bbs} \|\bbPsi\bbs\|_1 
    &\nonumber\\&
    \qquad
    \st \|\bbSigma\bbs + \bbM\hat{\bbp}\|_2 \leq \epsilon,
    ~(\bbe_1\otimes{\bf 1}_{O-1})^\top\bbs = 1,
    &\nonumber\\&
    \qquad \qquad \!
    \| \bbs - \hat{\bbs} \|_2 \leq C_s.
    &\label{e:l1_am_s_laststep}
&\end{alignat}
Thus, the convergence of proximal alternating minimization allows us to consider minimization with respect to $\bbs$ for both~\eqref{e:l0_vec_prob} and~\eqref{e:l1_vec_prob}.

We next consider when the solutions to~\eqref{e:l0_am_s_laststep} and~\eqref{e:l1_am_s_laststep} are equivalent.
We introduce a modification to~\eqref{e:l1_am_s_laststep} without the last constraint
\begin{alignat}{3}&
    \bar{\bbs} \in \argmin_{\bbs} \|\bbPsi\bbs\|_1 
    &\nonumber\\&
    \qquad
    \st \|\bbSigma\bbs + \bbM\hat{\bbp}\|_2 \leq \epsilon,
    ~(\bbe_1\otimes{\bf 1}_{O-1})^\top\bbs = 1,
    &\label{e:l1_am_s_mod}
&\end{alignat}
which may not have a unique solution.
By~\eqref{e:l0_l1_model}, $\bbs'$ is a feasible solution to~\eqref{e:l1_am_s_mod}.

By the proof of Theorem 1 in \cite{navarro2022joint} and Theorem 1 of \cite{zhang2016one}, if $\bbSigma_{\cdot,\ccalI}$ is full column rank and there exists a constant $\psi>0$ such that 
\begin{equation}\label{e:thm1_cond2}
    \|\bbPsi_{\ccalJ^c,\cdot}(\psi^{-2}\bbT+\bbPsi_{\ccalJ^c,\cdot}^\top\bbPsi_{\ccalJ^c,\cdot})^{-1}\bbPsi_{\ccalJ,\cdot}^\top\|_\infty<1,
\end{equation}
then we not only have that ${\bbs'}=\bar{\bbs}$, but $\bbs'$ is also the unique solution to~\eqref{e:l1_am_s_mod}.
These are exactly conditions \textit{1)} and \textit{2)} in the statement of Theorem 1.
Thus, we need only show that $\bar{\bbs} = \hat{\bbs}$.

Since~\eqref{e:l1_am_s_laststep} and~\eqref{e:l1_am_s_mod} share the first two constraints and $\|\hat{\bbs}-\bbs'\|_2 = \|\hat{\bbs} - \bar{\bbs}\|_2 \leq C_s$, $\hat{\bbs}$ and $\bar{\bbs}$ are both feasible solutions for~\eqref{e:l1_am_s_laststep} and~\eqref{e:l1_am_s_mod}.
Moreover, both problems have unique solutions, so $\hat{\bbs}=\bar{\bbs}=\bbs'$, as desired.

\section{Proof of Theorem 2}\label{B:thm2_proof}
To establish an upper bound on the estimation error of \eqref{e:l1_prob}, we first provide the following lemma necessary to determine an upper bound on the error of \eqref{e:l1_prob}.

\medskip

\noindent{\bf Lemma 1.}\textit{
% Under conditions of Theorem 2,
Under the following four conditions, 
    \bi[\it 1)] $K = o(\log O)$;
    \i[\it 2)] $R_1 \asymp R_2 \asymp \cdots \asymp R_K $;
    \i[\it 3)] $\log O = o(\min\{ R/(K^7(\log R)^2), (R/K^7)^{1/3} \})$; and
    \i[\it 4)] $\epsilon_R \geq CO\omega\sqrt{(K\log O)/R}$ for some constant $C>0$;
    \ei
with probability at least $1-e^{-C_1\log O}$ for some constant $C_1$ we have that 
\be
    \sum_{k=1}^K \left\|(\hbCo^{(k)} - \bbCo^{(k)})\bbSo^{*(k)}-\bbSo^{*(k)}(\hbCo^{(k)} - \bbCo^{(k)})\right\|_F^2 \leq \epsilon_R^2.
\ee
}

\medskip

\noindent{\bf Proof.} 
The proof of Lemma 1 follows from the proof of Claim 2 in \cite{navarro2022joint}.
$\hfill\square$
\medskip

Recall that $\bbs^*$ is the vectorization of the target GSO submatrices $\{\bbSo^{*(k)}\}_{k=1}^K$ as in \eqref{e:vecs_s}.
We show that $\{\bbs^*,\hat{\bbp}\}$ is a feasible solution to \eqref{e:l1_vec_prob}.
We demonstrate an upper bound on the commutativity of sample covariance submatrices and target subnetworks as
\begin{alignat}{3}&
    \bigg| \sum_{k=1}^K \|\hbCo^{(k)}\bbSo^{*(k)} - \bbSo^{*(k)}\hbCo^{(k)} + \hat{\bbP}^{(k)} - (\hat{\bbP}^{(k)})^\top\|_F^2 \bigg|^{\frac{1}{2}}
    &\nonumber\\&
    ~
    \leq \bigg| \sum_{k=1}^K \left\|(\hbCo^{(k)} - \bbCo^{(k)})\bbSo^{*(k)}-\bbSo^{*(k)}(\hbCo^{(k)} - \bbCo^{(k)})\right\|_F^2
    \bigg|^{\frac{1}{2}}
    &\nonumber\\&
    ~~
    + \bigg| \sum_{k=1}^K\left\|\big(\hat{\bbP}^{(k)}-(\hat{\bbP}^{(k)})^\top\big)-\big(\bbP^{*(k)}-(\bbP^{*(k)})^\top\big)\right\|_F^2
    \bigg|^{\frac{1}{2}}
    &\nonumber\\&
    ~
    \leq \epsilon_R + \alpha,
    \label{e:commut_uppbound}
&\end{alignat}
where we have used Lemma 1, the definition of $\alpha$, and the relationship in \eqref{e:commut}.
Because $\sum_{j=1}^O [\bbSo^{*(k)}]_{j1}=1$ by definition, \eqref{e:commut_uppbound} is equivalent to
\begin{equation}\label{e:commut_uppbound_vec}
    \|\bbSigma\bbs^*+\bbM\hat{\bbp}\|_2 \leq \epsilon_R+\alpha = \epsilon,
\end{equation} 
so $\{\bbs^*,\hat{\bbp}\}$ is a feasible solution to \eqref{e:l1_vec_prob}.

% Modification introduced for the introduction of \bar{\bbPhi}_r
We introduce a modification of \eqref{e:l1_vec_prob} to combine the constraints into one inequality.
Consider the following modified optimization problem that is parameterized by $r>0$ 
\begin{alignat}{3}&
    \{\hat{\bbs}_r,\hat{\bbp}_r\} = \argmin_{\{\bbs,\bbp\}} ~~
    \|\bbPsi\bbs\|_1 + 
    \sum_{g\in\ccalE} \eta_g' \|\bbp_g \|_2
    &\nonumber\\&
    \qquad \qquad \quad
    \st
    ~~
    \|\bar{\bbPhi}_r\bbs + \bar{\bbR}\bbp - \bar{\bbb}_r\|_2 \leq \epsilon,
    \label{e:param_prob}
&\end{alignat}
% \begin{alignat}{3}&
%     \{\hat{\bbs}_r,\hat{\bbp}_r\} = \argmin_{\{\bbs,\bbp\}} ~~
%     \|\bbPsi\bbs\|_1 + \sum_{k=1}^K \sum_{i=1}^O \gamma_k\|\bbp_{\ccalE^{(k,i)}}\|_2
%     &\nonumber\\&
%     \qquad \qquad \qquad \qquad \qquad
%     + \sum_{k<k'} \sum_{i=1}^O \eta_{k,k'} \|\bbp_{\ccalE^{(k,k',i)}}\|_2
%     &\nonumber\\&
%     \qquad \qquad \quad
%     \st
%     ~~
%     \|\bar{\bbPhi}_r\bbs + \bar{\bbR}\bbp - \bar{\bbb}_r\|_2 \leq \epsilon,
%     \label{e:param_prob}
% &\end{alignat}
where $\bar{\bbPhi}_r = [ \bbSigma^\top,r(\bbe_1\otimes{\bf 1}_{O-1}) ]^\top$, $\bar{\bbR} = [\bbM^\top,{\bf 0}_{KO^2}]^\top$, and $\bar{\bbb}_r = [{\bf 0}_{KO(O-1)/2}^\top,r]^\top$.
The parameter $r$ determines the strictness of the second constraint in \eqref{e:l1_vec_prob} such that when $r\rightarrow\infty$, we have that $\hat{\bbs}_r \rightarrow \hat{\bbs}$.
Note that since $(\bbe_1\otimes{\bf 1}_{O-1})^\top \hat{\bbs}=1$ and $(\bbe_1\otimes{\bf 1}_{O-1})^\top {\bbs^*}=1$, then by \eqref{e:commut_uppbound_vec} and the definition of $\{\hat{\bbs},\hat{\bbp}\}$, we have that $\{\hat{\bbs},\hat{\bbp}\}$ and $\{{\bbs^*},\hat{\bbp}\}$ are feasible solutions of \eqref{e:param_prob} for every $r>0$.

We next provide an upper bound on the difference between $\hat{\bbs}$ and $\bbs^*$ following the proof of Claim 1 in \cite{navarro2022joint}.
Recall that we define $\ccalK$ as $\mathrm{supp}(\bbPsi\bbs^*)$.
First, note that as in the proof of Claim 1 of \cite{navarro2022joint}, we have that when $\bbSigma$ is full column rank, then so is $\bar{\bbPhi}_r$, which guarantees the existence of a dual certificate $\bby = \bbI_{\ccalK,\cdot}^\top\mathrm{sign}(\bbPsi_{\ccalK,\cdot}\bbs^*)$, where $\bbPsi^\top\bby = \bar{\bbPhi}_r^\top\bar{\bbPhi}_r(\bar{\bbPhi}_r^\top\bar{\bbPhi}_r)^{-1}\bbPsi^\top\bbI_{\ccalK,\cdot}^\top\mathrm{sign}(\bbPsi_{\ccalK,\cdot}\bbs^*)\in\mathrm{Im}(\bar{\bbPhi}_r^\top)$, $\bby_{\ccalK}=\mathrm{sign}(\bbPsi_{\ccalK,\cdot}\bbs^*)$, $\|\bby_{\ccalK^c}\|_\infty<1$, and $\|\bbPsi\bbs^*\|_1=\bby^\top\bbPsi\bbs^*$.

Consider the following inequality
\begin{alignat}{3}
    \| \bbPsi\bbs^*-\bbPsi\hat{\bbs} \|_1
    \leq
    \| \bbPsi\hat{\bbs}-\bbu \|_1 + \| \bbPsi\bbs^*-\bbu \|_1,
    \label{e:claim1_psidiff}
\end{alignat}
where $\bbu\in\mathbb{R}^{KO(O-1)/2}$ such that $\mathrm{supp}(\bbu)\subseteq\ccalK$.
We derive an upper bound for the second term on the right-hand side of \eqref{e:claim1_psidiff} as
\begin{alignat}{3}&
    \| \bbPsi\bbs^*-\bbu \|_1
    & \leq &
    \sqrt{|\ccalK|}\| \bbPsi\bbs^*-\bbu \|_2
    &\nonumber\\&
    & \leq &
    \sqrt{|\ccalK|}\| \bbPsi\bbs^*-\bbPsi\hat{\bbs} \|_2 + \sqrt{|\ccalK|}\| \bbPsi\hat{\bbs}-\bbu \|_1
    &\nonumber\\&
    & \leq &
    \sqrt{|\ccalK|}\sigma_{\mathrm{max}}(\bbPsi)\| \bbs^*-\hat{\bbs} \|_2 
    &\nonumber\\&
    && \qquad
    + \sqrt{|\ccalK|}\| \bbPsi\hat{\bbs}-\bbu \|_1
    &\nonumber\\&
    & \leq &
    \frac{\sqrt{|\ccalK|}\sigma_{\mathrm{max}}(\bbPsi)}{\sigma_{\mathrm{min}}(\bar{\bbPhi}_r)}
    \| \bar{\bbPhi}_r(\bbs^*-\hat{\bbs}) \|_2 
    &\nonumber\\&
    & & \qquad
    + \sqrt{|\ccalK|}\| \bbPsi\hat{\bbs}-\bbu \|_1.
    \label{e:claim1_term2}
&\end{alignat}
For the first term on the right-hand side of \eqref{e:claim1_psidiff}, we have that 
\begin{alignat}{3}&
    \xi &~:=~& \min_{\bbu:\mathrm{supp}(\bbu)\subseteq\ccalK} \| \bbPsi\hat{\bbs}-\bbu \|_1
    &\nonumber\\&
    &~=~& \max_{\bbv} \min_{\bbu} \| \bbPsi\hat{\bbs}-\bbu \|_1 
    &\label{e:claim1_term1_pt1a}\\&
    && \quad + \bbv^\top\bbI_{\ccalK^c,\cdot}(\bbu-\bbPsi\hat{\bbs}) 
    + \bbv^\top\bbI_{\ccalK^c,\cdot}\bbPsi\hat{\bbs}
    &\nonumber\\&
    &~=~& \max_{\bbw:\mathrm{supp}(\bbw)\subseteq\ccalK^c} \min_{\bbu} \| \bbPsi\hat{\bbs}-\bbu \|_1
    &\nonumber\\&
    && \quad + \bbw^\top(\bbu-\bbPsi\hat{\bbs}) 
    + \bbw^\top\bbPsi\hat{\bbs},
    \nonumber
&\end{alignat}
where \eqref{e:claim1_term1_pt1a} results from the Lagrangian of $\xi$ and duality theory.
Given the dual certificate $\bby$, we have that
\begin{alignat}{3}&
    \xi &~=~& \max_{\substack{\bbw:\mathrm{supp}(\bbw)\subseteq\ccalK^c, \\ \|\bbw\|_\infty\leq 1}} (\bby+\bbw)^\top\bbPsi\hat{\bbs} 
    - \bby^\top\bbPsi\hat{\bbs}
    &\nonumber\\&
    &~\leq~& \|\bbPsi\hat{\bbs}\|_1 - \bby^\top\bbPsi\hat{\bbs} + \bby^\top\bbPsi{\bbs^*} - \|\bbPsi{\bbs^*}\|_1
    &\nonumber\\&
    &~\leq~& \bby^\top\bbPsi(\bbs^*-\hat{\bbs}),
    \label{e:claim1_term1_pt2}
&\end{alignat}
where the final inequality is due to the optimality of $\{\hat{\bbs},\hat{\bbp}\}$ and the feasibility of $\{\bbs^*,\hat{\bbp}\}$ for \eqref{e:l1_vec_prob}.
Lastly, since $\bbPsi^\top\bby=\bar{\bbPhi}_r^\top\bar{\bbPhi}_r(\bar{\bbPhi}_r^\top\bar{\bbPhi}_r)^{-1}\bbPsi^\top\bbI_{\ccalK,\cdot}^\top\mathrm{sign}(\bbPsi_{\ccalK,\cdot}\bbs^*)$, we have that
\begin{alignat}{3}&
    \bby^\top\bbPsi(\bbs^*-\hat{\bbs})
    &\nonumber\\&
    \qquad \leq
    \mathrm{sign}(\bbPsi_{\ccalK,\cdot}\bbs^*)^\top\bbI_{\ccalK,\cdot}\bbPsi(\bar{\bbPhi}_r^\top\bar{\bbPhi}_r)^{-1}\bar{\bbPhi}_r^\top\bar{\bbPhi}_r(\bbs^*-\hat{\bbs})
    &\nonumber\\&
    \qquad \leq
    \frac{\sqrt{|\ccalK|}\sigma_{\mathrm{max}}(\bbPsi)}{\sigma_{\mathrm{min}}(\bar{\bbPhi}_r)}
    \| \bar{\bbPhi}_r(\bbs^*-\hat{\bbs}) \|_2,
    \label{e:claim1_term1_pt3}
&\end{alignat}
where the second inequality results from the fact that every positive scalar and its $\ell_2$ norm are equal.
We may substitute \eqref{e:claim1_term2} and \eqref{e:claim1_term1_pt3} into \eqref{e:claim1_psidiff} and the fact that $\bbPsi$ is full column rank to obtain
\begin{alignat}{3}&
    \|\bbs^* - \hat{\bbs}\|_1 \leq \tau_r\|\bar{\bbPhi}_r(\bbs^*-\hat{\bbs})\|_2,
    \nonumber
&\end{alignat}
where 
\begin{equation}\label{e:tau_r}
\tau_r = \frac{\sqrt{|\ccalK|}\sigma_{\mathrm{max}}(\bbPsi)\|\bbPsi^{\dagger}\|_1}{\sigma_{\mathrm{min}}(\bar{\bbPhi}_r)}(2+\sqrt{|\ccalK|}).
\end{equation}

As $r\rightarrow\infty$, we have that 
\begin{alignat}{3}&
    \|\bbs^* - \hat{\bbs}\|_1 
    &~\leq~&
    \lim_{r\rightarrow\infty} \tau_r\|\bar{\bbPhi}_r(\bbs^*-\hat{\bbs})\|_2
    &\nonumber\\&
    &~\leq~&
    2\lim_{r\rightarrow\infty}\tau_r(\epsilon_R + \alpha),
    \nonumber
&\end{alignat}
where by the feasibility of $\{\hat{\bbs},\hat{\bbp}\}$ and $\{\bbs^*,\hat{\bbp}\}$ for every $r>0$, we have that \begin{alignat}{3}&
    \|\bar{\bbPhi}_r(\bbs^*-\hat{\bbs})\|_2
    &~\leq~&
    \|\bar{\bbPhi}_r\bbs^*+\bar{\bbR}\hat{\bbp}-\bar{\bbb}_r\|_2
    &\nonumber\\&
    && \quad +
    \|\bar{\bbPhi}_r\hat{\bbs}+\bar{\bbR}\hat{\bbp}-\bar{\bbb}_r\|_2
    &\nonumber\\&
    &~\leq~& 
    2(\epsilon_R + \alpha).
&\end{alignat}

Finally, we return to the equivalent matrix formulation as
\begin{alignat}{3}&
    \sum_{k=1}^K \| \hbSo^{(k)} - \bbSo^{*(k)} \|_1 \leq 4\tau_r(\epsilon_R + \alpha).
&\end{alignat}
% By the end of the proof of Theorem 2 in \cite{navarro2022joint}, we have that \blue{$\lim_{r\rightarrow\infty}8\tau_r\leq\tau$}, proving the first inequality in Theorem~2 of this work as desired.
By the end of the proof of Theorem 2 in \cite{navarro2022joint}, we have that $\lim_{r\rightarrow\infty}4\tau_r\leq\tau$, as desired.

\section{Proof of Corollary 1}\label{C:coroll_proof}

Consider the following optimization problem
\begin{alignat}{3}&
    \min_{ \{\bbSo^{(k)}\}_{k=1}^K } ~
    \sum_{k=1}^K \alpha_k \| \bbSo^{(k)} \|_1
     + \sum_{k<k'} \beta_{k,k'} \| \bbSo^{(k)}-\bbSo^{(k')} \|_1
    &\nonumber\\&
    \st ~\!\!
    \textstyle\sum_{k=1}^K \| \hbCo^{(k)}\bbSo^{(k)} - \bbSo^{(k)}\hbCo^{(k)} \|_F^2 \leq \epsilon_R^2,
    &\nonumber\\&
    \qquad ~\!\!
    \bbSo^{(k)}=(\bbSo^{(k)})^\top, ~ 
    \mathrm{diag}(\bbSo^{(k)}) = {\bf 0}, ~ 
    \forall k=1,\dots, K,
    &\nonumber\\&
    \qquad\!\!
    \textstyle\sum_{j} [\bbSo^{(1)}]_{j1} = 1,
    \label{e:naive_prob}
&\end{alignat}
whose solution is equivalent to the naive solution $\{\tbSo^{(k)}\}_{k=1}^K$ described in the statement of Corollary 1.
Similarly to \eqref{e:l1_prob}, we can define a vectorized version of \eqref{e:naive_prob} as
\begin{alignat}{3}
    &\tilde{\bbs} = \argmin_{\bbs }
    \|\bbPsi\bbs\|_1
    ~
    \st
    ~
    \|\bbSigma\bbs\|_2 \leq \epsilon_R,
    ~(\bbe_1\otimes{\bf 1}_{O-1})^\top\bbs = 1,
    \label{e:naive_vec_prob}
&\end{alignat}
and a version parameterized by $r>0$ as 
\begin{alignat}{3}
    &\tilde{\bbs}_r = \argmin_{\bbs }
    \|\bbPsi\bbs\|_1
    ~
    \st
    ~
    \|\bar{\bbPhi}_r\bbs-\bar{\bbb}_r\|_2 \leq \epsilon_R,
    \label{e:naive_param_prob}
&\end{alignat}
where $\bar{\bbPhi}_r$ and $\bar{\bbb}_r$ are defined as for \eqref{e:param_prob} and $\lim_{r\rightarrow\infty}\tilde{\bbs}_r=\tilde{\bbs}$.

We provide the following upper bound via \eqref{e:commut}  
\begin{alignat}{3}&
    \bigg| \sum_{k=1}^K 
    \|\hbCo^{(k)}\bbSo^{*(k)} - \bbSo^{*(k)}\hbCo^{(k)}\|_F^2 \bigg|^{\frac{1}{2}}
    &\nonumber\\&
    ~~
    \leq \bigg| \sum_{k=1}^K \left\|(\hbCo^{(k)} - \bbCo^{(k)})\bbSo^{*(k)}-\bbSo^{*(k)}(\hbCo^{(k)} - \bbCo^{(k)})\right\|_F^2
    \bigg|^{\frac{1}{2}}
    &\nonumber\\&
    ~~
    + \bigg| \sum_{k=1}^K\left\|\bbP^{*(k)}-(\bbP^{*(k)})^\top\right\|_F^2
    \bigg|^{\frac{1}{2}}
    &\nonumber\\&
    ~~\leq \epsilon_R + \delta,
    \nonumber
&\end{alignat}
and similarly to Theorem 2, we apply Lemma 1 to get
\be
    \|\bar{\bbPhi}_r\bbs^*-\bar{\bbb}_r\|_2 \leq \epsilon_R + \delta,
\ee
where $\bbs^*$ may not be a feasible solution to \eqref{e:naive_param_prob}.
However, by the triangle inequality and the optimality of $\tilde{\bbs}_r$, there exists $\rho\in[0,1]$ such that 
\begin{equation}\label{e:infeas_strue}
    \|\bbPsi\tilde{\bbs}_r\|_1 - \|\bbPsi{\bbs}^{*}\|_1 \leq \rho\|\bbPsi\tilde{\bbs}_r-\bbPsi{\bbs}^{*}\|_1.
\end{equation}
In particular, let $\rho = \max\{0, (\|\bbPsi\tilde{\bbs}_r\|_1 - \|\bbPsi{\bbs}^{*}\|_1)/\|\bbPsi\tilde{\bbs}_r-\bbPsi{\bbs}^{*}\|_1\}$, where $\rho=0$ when $\bbs^*$ is a feasible solution to \eqref{e:naive_param_prob}, but otherwise, it may be possible that $\rho\in(0,1]$.
Furthermore, since $(\bbe_1\otimes{\bf 1}_{O-1})^\top \tilde{\bbs}=1$, then $\tilde{\bbs}$ is a feasible solution to \eqref{e:naive_param_prob} for every $r>0$.

We then can introduce a similar inequality to \eqref{e:claim1_psidiff} as 
\begin{alignat}{3}
    \| \bbPsi\bbs^*-\bbPsi\tilde{\bbs} \|_1
    \leq
    \| \bbPsi\tilde{\bbs}-\tilde{\bbu} \|_1 + \| \bbPsi\bbs^*-\tilde{\bbu} \|_1,
    \label{e:corrol_psidiff}
\end{alignat}
where $\tilde{\bbu}\in\mathbb{R}^{KO(O-1)/2}$ such that $\mathrm{supp}(\tilde{\bbu})\subseteq\ccalK$.
The upper bound for the second term of the right-hand side of \eqref{e:corrol_psidiff} can be found analogously to \eqref{e:claim1_term2}, where we have 
\begin{alignat}{3}&
    \| \bbPsi\bbs^*-\tilde{\bbu} \|_1
    & \leq &
    \frac{\sqrt{|\ccalK|}\sigma_{\mathrm{max}}(\bbPsi)}{\sigma_{\mathrm{min}}(\bar{\bbPhi}_r)}
    \| \bar{\bbPhi}_r(\bbs^*-\tilde{\bbs}_r) \|_2 
    &\nonumber\\&
    & & \qquad
    + \sqrt{|\ccalK|}\| \bbPsi\tilde{\bbs}_r-\tilde{\bbu} \|_1.
    \label{e:claim1_term2_pt2}
&\end{alignat}
Similarly to \eqref{e:claim1_term1_pt2} in the proof of Theorem 2, we can upper bound the first term as
\begin{alignat}{3}&
    \tilde{\xi} &~:=~& \min_{\tilde{\bbu}:\mathrm{supp}(\tilde{\bbu})\subseteq\ccalK} \| \bbPsi\tilde{\bbs}-\tilde{\bbu} \|_1
    &\nonumber\\&
    &~\leq~&
    \|\bbPsi\tilde{\bbs}\|_1 - \bby^\top\bbPsi\tilde{\bbs} + \bby^\top\bbPsi{\bbs^*} - \|\bbPsi{\bbs^*}\|_1
    &\nonumber\\&
    &~\leq~& \bby^\top\bbPsi(\bbs^*-\tilde{\bbs}) + \rho\|\bbPsi(\bbs^*-\tilde{\bbs})\|_1,
    \label{e:corrol_term1}
&\end{alignat}
where we account for the possible infeasibility of $\bbs^*$ with \eqref{e:infeas_strue}.
We may combine \eqref{e:corrol_term1}, and \eqref{e:claim1_term2_pt2} to obtain
\begin{alignat}{3}&
    \|\tilde{\bbs}-\bbs^*\|_1 \leq (\tau_r + \tau_r')(2\epsilon_R+\delta),
&\end{alignat}
where $\tau_r$ is defined in \eqref{e:tau_r} and we let 
\be
    \tau'_r := \frac{\rho KO(O-1) (1+\sqrt{|\ccalK|})\sigma_{\mathrm{max}}(\bbPsi)\|\bbPsi^\dagger\|_1}{2\sigma_{\mathrm{min}}(\bar{\bbPhi}_r)}.
\ee

As with the proof of Theorem 2, we have that for $r\rightarrow\infty$,
\begin{alignat}{3}&
    \sum_{k=1}^K \| \tbSo^{(k)} - \bbSo^{*(k)} \|_1 \leq (\tau+\tau')(\epsilon_R + \tfrac{1}{2}\delta),
&\end{alignat}
as desired.

Moreover, the bound \eqref{e:corol_pt2} is equivalent to the following inequality
\be
    \alpha^2 \leq \left(\frac{\tau'}{\tau}\right)^2\epsilon_R^2 + \left(\frac{\tau+\tau'}{2\tau}\right)^2\delta^2,
\ee
which is a sufficient condition for the upper bound in \eqref{e:thm2} to be less than the upper bound in \eqref{e:corol_p1}.

\section{Proof of Theorem 3}\label{D:thm3_proof}

The proof of Theorem 3 is inspired by that of~\cite[Theorem 1]{percival2012theoretical}.
Analogous to how we obtain the inequality~\eqref{e:commut_uppbound_vec}, we apply Lemma 1, the definition of $\alpha$, and~\eqref{e:commut} to conclude that
\begin{equation*}
    \| \bbSigma\bbs^* + \bbM \bbp^* \|_2 \leq \epsilon_R.
\end{equation*}
Then, by~\eqref{e:commut} and the convexity of~\eqref{e:l1_vec_prob}, we represent $\hbp$ as
\begin{equation}\label{e:p_subprob}
    \hbp \in 
    \argmin_\bbp
    \mu \|\bbSigma\hbs + \bbM\bbp\|_2^2
    + \sum_{g\in\ccalE} \eta_g' \|\bbp_g\|_2,
\end{equation}
where by optimization theory there exists a constant $\mu>0$ such that~\eqref{e:p_subprob} is equivalent to
\begin{equation*}
    \hbp \in \argmin_\bbp
    \sum_{g\in\ccalE} \eta_g' \|\bbp_g\|_2
    ~\st~ \|\bbSigma\hbs + \bbM\bbp\|_2 \leq \epsilon_R + \alpha.
\end{equation*}
We first aim to bound the error of $\hbp$.
To this end, consider
\begin{alignat}{3}&
    \eta_\mathrm{min} \sum_{g\in\ccalE} \|\hbp_g - \bbp^*_g\|_2
    % &~\leq~&
    &\nonumber\\&
    \qquad \leq
    \sum_{g\in\ccalE} \eta_g' \|\hbp_g\|_2
    + \sum_{g\in\ccalE} \eta_g' \|\bbp_g^*\|_2
    &\nonumber\\&
    \qquad \leq
    % &~\leq~&
    2\sum_{g\in\ccalE} \eta_g' \|\bbp_g^*\|_2
    + \mu \|\bbSigma\hbs + \bbM\bbp^*\|_2^2
    - \mu \|\bbSigma\hbs + \bbM\hbp\|_2^2
    \nonumber
&\end{alignat}
since $\hbp$ minimizes~\eqref{e:p_subprob}.
Then, we have that
\begin{alignat}{3}&
    \eta_\mathrm{min} \sum_{g\in\ccalE} \|\hbp_g - \bbp^*_g\|_2
    &\nonumber\\&
    \qquad \leq
    2\sum_{g\in\ccalE} \eta_g' \|\bbp_g^*\|_2
    + 2\mu (\bbSigma\hbs + \bbM\bbp^*)^\top \bbM (\hbp-\bbp^*)
    &\nonumber\\&
    \qquad \leq
    2\sum_{g\in\ccalE} \eta_g' \|\bbp_g^*\|_2
    + 4\mu (\epsilon_R + \sigma\|\hbs - \bbs^*\|_1) \|\hbp-\bbp^*\|_2.
    \nonumber
&\end{alignat}
Furthermore, by applying the Cauchy-Schwartz inequality to the $\ell_2$ and $\ell_1$ norms, we obtain the following inequality for the group lasso penalties
\begin{equation*}
    \|\hbp-\bbp^*\|_2
    \leq
    \frac{2\sqrt{O}}{K}
    \sum_{g\in\ccalE} \|\hbp_g-\bbp^*_g\|_2,
\end{equation*}
so we have that
\begin{alignat}{3}&
    \|\hbp-\bbp^*\|_2
    \leq
    \frac{ 4 \sqrt{O} \sum_{g\in\ccalE} \eta_g' \|\bbp_g^*\|_2 }
         { K\eta_\mathrm{min} - 8\mu\sqrt{O} (\epsilon_R + \sigma\|\hbs-\bbs^*\|_1) }.
    \label{e:perr_s}
&\end{alignat}

We rewrite the error bound for $\{\hbSo^{(k)}\}_{k=1}^K$ in Theorem 2 in vectorized form, and by~\eqref{e:perr_s} we have
\begin{alignat}{3}&
    \|\hbs - \bbs^*\|_1
    &~\leq~&
    2\tau (\epsilon_R + \alpha)
    &\nonumber\\&
    &~\leq~&
    2\tau (\epsilon_R + 2\|\hbp-\bbp^*\|_2)
    &\nonumber\\&
    &~\leq~&
    2\tau\epsilon_R + 
    \frac{ 16 \tau \sqrt{O} \sum_{g\in\ccalE} \eta_g' \|\bbp_g^*\|_2 }
         { K\eta_\mathrm{min} - 8\mu\sqrt{O} (\epsilon_R + \sigma\|\hbs-\bbs^*\|_1) }.
    \nonumber
&\end{alignat}
This inequality is quadratic in $\|\hbs-\bbs^*\|_1$.
Solving it for $\|\hbs-\bbs^*\|_1$ yields the following bound
\begin{alignat}{3}&
    \|\hbs-\bbs^*\|_1
    &~\leq~&
    2\tau \epsilon_R + \sqrt{\frac{2\tau \sum_{g\in\ccalE} \eta_g' \|\bbp^*_g\|_2}{\mu\sigma}},
    \nonumber
&\end{alignat}
which is equivalent to the error bound~\eqref{e:thm3}.

In addition to providing the error bound for $\hbs$, assumptions~\eqref{e:etamin_lowbnd} and~\eqref{e:etamax_uppbnd} also ensure that there exist parameters $\{\eta_g'\}_{g\in\ccalE}$ such that the error bound~\eqref{e:thm3} is valid.
In particular, we have that
\begin{alignat}{3}&
    8\mu\sqrt{O}(\epsilon_R + \sigma\|\hbs - \bbs^*\|_1)
    &\nonumber\\&
    \qquad\quad \leq
    8\mu\sqrt{O} (\epsilon_R (2\tau\sigma+1) + 
    2\tau\sigma\kappa )
    &\nonumber\\&
    \qquad\quad \leq
    8\mu\epsilon_R\sqrt{O} ( 2\tau\sigma+1) + 8\eta_\mathrm{max}^{1/2} \sqrt{2O\mu\tau\sigma\sum_{g\in\ccalE} \|\bbp_g^*\|_2}
    &\nonumber\\&
    \qquad\quad <
    K\eta_\mathrm{min},
    \nonumber
&\end{alignat}
where the final inequality results from~\eqref{e:etamax_uppbnd}, so the denominator in~\eqref{e:perr_s} is strictly positive.
Then, by~\eqref{e:etamin_lowbnd} there exist parameters $\{\eta_g'\}_{g\in\ccalE}$ satisfying~\eqref{e:etamax_uppbnd} such that $\eta_\mathrm{min} \leq \eta_\mathrm{max}$.
Thus, the error bound~\eqref{e:thm3} holds for a valid set of parameters that follow the given assumptions.

We next prove the conditions under which we can guarantee that~\eqref{e:corol_pt2} holds, that is, the error bound of $\{\hbSo^{(k)}\}_{k=1}^K$ is lower than that of $\{\tbSo^{(k)}\}_{k=1}^K$.
First, we show that conditions~\eqref{e:grouplasso_uppbnd} and~\eqref{e:delta_lowbnd} guarantee that the lower bound in~\eqref{e:etamin_lowbnd} is strictly lower than the upper bound for $\eta_\mathrm{max}$ in~\eqref{e:etamax_thm3}.
To see this, we rewrite~\eqref{e:grouplasso_uppbnd} as
\begin{equation*}
    128\frac{O}{K^2} \mu\tau\sigma \left( \sum_{g\in\ccalE} \|\bbp_g^*\|_2 \right)^2
    <
    2\mu\sigma \frac{(\tau'\epsilon_R)^2}{\tau},
\end{equation*}
so the first term in the right-hand side of~\eqref{e:etamin_lowbnd} is strictly lower than the first term in the right-hand side of~\eqref{e:etamax_thm3}.

Second, we consider~\eqref{e:delta_lowbnd}.
By~\eqref{e:grouplasso_uppbnd} and~\eqref{e:delta_lowbnd}, we have that
\begin{alignat}{3}&
    16\frac{\sqrt{O}}{K}\mu\epsilon_R (2\tau\sigma+1)\cdot\frac{\tau\sum_{g\in\ccalE}\|\bbp_g^*\|_2}{2\mu\sigma}
    &~<~&
    \frac{\tau'\epsilon_R^2 (2\tau\sigma+1)}{\sigma}
    &\nonumber\\&
    &~<~&
    {\textstyle\frac{1}{4}} (\tau+\tau')^2 \delta^2,
    \nonumber
&\end{alignat}
so the second term of~\eqref{e:etamin_lowbnd} is strictly less than the second term of~\eqref{e:etamax_thm3}.
We thus guarantee that there exist parameters $\{\eta_g'\}_{g\in\ccalE}$ that satisfy $\eta_\mathrm{min}\leq\eta_\mathrm{max}$ and assumptions~\eqref{e:etamin_lowbnd},~\eqref{e:etamax_uppbnd}, and~\eqref{e:etamax_thm3}.

Finally, we show when~\eqref{e:etamax_thm3} implies~\eqref{e:corol_pt2}.
We rewrite~\eqref{e:etamax_thm3} to get
\begin{alignat}{3}&
    \kappa^2
    &~\leq~&
    \frac{\eta_\mathrm{max}\sum_{g\in\ccalE}\|\bbp_g^*\|_2}{2\tau\mu\sigma}
    &\nonumber\\&
    &~\leq~&
    \frac{(\tau'\epsilon_R)^2}{\tau^2}
    + \frac{(\tau+\tau')^2\delta^2}{4\tau^2},
&\end{alignat}
which is sufficient for the upper bound $\tau(\epsilon_R + \kappa)$ to be less than the error bound $\tau (\epsilon_R + {\textstyle\frac{1}{2}}\delta)$ in Corollary 1.
If $\alpha\leq\kappa$, then we can guarantee that~\eqref{e:corol_pt2} holds, but for $\alpha > \kappa$ we achieve a stronger result comparing the error bounds for $\hbSo^{(k)}$ and $\tbSo^{(k)}$.

}

% references section
\bibliographystyle{IEEEtran}
\bibliography{biblio}

\end{document}

%% file: hidden_nodes_defs.tex
\def \bbCo {\bbC_{{\scriptscriptstyle \ccalO}}}
\def \bbCh {\bbC_{{\scriptscriptstyle \ccalH}}}
\def \bbCoh {\bbC_{{\scriptscriptstyle \ccalO\ccalH}}}
\def \bbCho {\bbC_{{\scriptscriptstyle \ccalH\ccalO}}}

\def \hbCo {\hbC_{{\scriptscriptstyle \ccalO}}}

\def \bbXo {\bbX_{{\scriptscriptstyle \ccalO}}}
\def \bbXh {\bbX_{{\scriptscriptstyle \ccalH}}}

\def \bbSo {\bbS_{{\scriptscriptstyle \ccalO}}}
\def \bbSh {\bbS_{{\scriptscriptstyle \ccalH}}}
\def \bbSoh {\bbS_{{\scriptscriptstyle \ccalO\ccalH}}}
\def \bbSho {\bbS_{{\scriptscriptstyle \ccalH\ccalO}}}

\def \hbSo {\hbS_{{\scriptscriptstyle \ccalO}}}

\def \bbSo {\bbS_{{\scriptscriptstyle \ccalO}}}
\def \bbSh {\bbS_{{\scriptscriptstyle \ccalH}}}
\def \bbSoh {\bbS_{{\scriptscriptstyle \ccalO\ccalH}}}
\def \bbSho {\bbS_{{\scriptscriptstyle \ccalH\ccalO}}}

\def \tbSo {\tilde{\bbS}_{{\scriptscriptstyle \ccalO}}}

%% file: figs/thm2.tex
\begin{tikzpicture}[baseline,scale=.93,trim axis left, trim axis right]

\pgfplotstableread{data/thm2_scaled.csv}\errtable

% \begin{axis}[
%     % scale only axis,
%     xlabel={(b) Edge weights in $\bbSoh^{*(k)}$},
%     xmin=0,
%     xmax=1,
%     ylabel={Normalized $\ell_1$ error},
%     ymin=0,
%     ymax=1.7,
%     grid style=densely dashed,
%     grid=both,
%     legend style={
%         at={(1,1)},
%         anchor=north east},
%     legend columns=1,
%     width=185,
%     height=160
%     ]
    
%     % \addlegendimage{}
%     \addplot[orange!75!white, mark=+, densely dotted] table [x=xaxis, y=Fphat] {\errtable};
%     \addplot[orange, mark=o, solid] table [x=xaxis, y=alpha] {\errtable};
%     \addplot[orange!65!white, mark=x, dashed] table [x=xaxis, y=delta] {\errtable};
%     \addplot[blue, mark=o, solid] table [x=xaxis, y=L1Shat] {\errtable};
%     \addplot[blue!65!white, mark=x, dashed] table [x=xaxis, y=L1Stil] {\errtable};

%     \legend{}
%     \addlegendentry{Joint-H}
%     \addlegendentry{Joint-nH}
%     \addlegendentry{Error of $\hat{\bbP}^{(k)}$}
%     \addlegendentry{$\alpha/C$}
%     \addlegendentry{$\delta/C$}
% \end{axis}

\begin{axis}[
    % scale only axis,
    xlabel={},
    xtick={},
    xmin=0,
    xmax=1,
    axis y line*=right,
    ymin=.8,
    ymax=.95,
    grid style=densely dashed,
    grid=both,
    legend style={
        at={(1,1)},
        anchor=north east},
    legend columns=1,
    width=185,
    height=160,
    xtick pos=left,
    ytick pos=right
    ]
    
    \addlegendimage{empty legend}
    % \addplot[orange!75!white, mark=+, densely dotted] table [x=xaxis, y=Fphat] {\errtable};
    \addplot[orange, mark=o, solid] table [x=xaxis, y=alpha] {\errtable};
    \addplot[orange!65!white, mark=x, dashed] table [x=xaxis, y=delta] {\errtable};
    
    \addlegendentry{\hspace{-.5cm}\textbf{Right axis}}
    % \addlegendentry{Error of $\hat{\bbP}^{(k)}$}
    \addlegendentry{$\bar{\alpha}/C$}
    \addlegendentry{$\bar{\delta}/C$}
\end{axis}

\begin{axis}[
    % scale only axis,
    xlabel={(a) Edge weights in $\bbSoh^{*(k)}$},
    xmin=0,
    xmax=1,
    axis y line*=left,
    ylabel={Normalized $\ell_1$ error},
    ymin=0,
    ymax=1.55,
    % grid style=densely dashed,
    % grid=both,
    legend style={
        at={(0,1)},
        anchor=north west},
    legend columns=1,
    width=185,
    height=160,
    xtick pos=left,
    ytick pos=left
    ]
    
    \addlegendimage{empty legend}
    \addplot[blue, mark=o, solid] table [x=xaxis, y=L1Shat] {\errtable};
    \addplot[blue!65!white, mark=x, dashed] table [x=xaxis, y=L1Stil] {\errtable};

    % \legend{Joint-H,Joint-nH}
    \addlegendentry{\hspace{-.4cm}\textbf{Left axis}}
    \addlegendentry{JH-GSR}
    \addlegendentry{J-GSR}
    % \addlegendentry{Joint-H}
    % \addlegendentry{Joint-nH}
\end{axis}

\end{tikzpicture}

%% file: figs/n_graphs.tex
\begin{tikzpicture}[baseline,scale=.93,trim axis left, trim axis right]

\pgfplotstableread{data/n_graphs.csv}\errtable

\begin{axis}[
    xlabel={(b) Number of graphs},
    xmin=1,
    xmax=6,
    xtick={1,2,...,6},
    ylabel={$\sum_{k=1}^K \mathrm{nerr}(\bbSo^{*(k)},\hbSo^{(k)})/K$},
    ymin=0,
    ymax=0.4,
    ytick={0,.1,...,.4},
    grid style=densely dashed,
    grid=both,
    legend style={
        at={(0,0)},
        anchor=south west},
    legend columns=2,
    width=185,
    height=160
    ]

    \addplot[blue, mark=o, solid] table [x=x, y=Joint-H] {\errtable}; 
    \addplot[blue!70!white, mark=o, densely dotted] table [x=x, y=Sep-H] {\errtable};
    \addplot[red, mark=x, solid] table [x=x, y=Joint-nH] {\errtable};
    \addplot[red!70!white, mark=x, densely dotted] table [x=x, y=Sep-nH] {\errtable};
    \legend{JH-GSR,SH-GSR,J-GSR,S-GSR}
    % \legend{Joint-H, Sep-H, Joint-nH, Sep-nH}

    % Could I read the first column?
    %\def\models {Unc, Tr, Sq, Heat, Sqrt, Poly};
    
    % Probably faster to just copy and paste this time...
    % try using cycle list for colors and markers!
    % \foreach \model in \models{
    %     \addplot table [x=xaxis, y=\model] {\errtable};
    %     \addlegendentryexpanded{\model}
    % }
    
\end{axis}
\end{tikzpicture}

%% file: figs/hidden_nodes.tex
\begin{tikzpicture}[baseline,scale=.93,trim axis left, trim axis right]

\pgfplotstableread{data/hidden.csv}\errtable

\begin{axis}[
    xlabel={(c) Hidden nodes},
    xmin=1,
    xmax=5,
    ylabel={$\sum_{k=1}^K \mathrm{nerr}(\bbSo^{*(k)},\hbSo^{(k)})/K$},
    ymin=0.1,
    ymax=.8,
    grid style=densely dashed,
    grid=both,
    legend style={
        at={(1,0)},
        anchor=south east},
    legend columns=3,
    transpose legend,
    width=185,
    height=160
    ]
    
    \addplot[blue, mark=o, solid] table [x=xaxis, y=PGLK2] {\errtable};
    \addplot[red, mark=x, solid] table [x=xaxis, y=PNNK2] {\errtable};
    \addplot[orange, mark=+, solid] table [x=xaxis, y=No-HK2] {\errtable};
    
    \addplot[blue!65!white, mark=o, densely dotted] table [x=xaxis, y=PGLK6] {\errtable};
    \addplot[red!65!white, mark=x, densely dotted] table [x=xaxis, y=PNNK6] {\errtable};
    \addplot[orange!65!white, mark=+, densely dotted] table [x=xaxis, y=No-HK6] {\errtable};

    % \legend{PGL-2, PNN-2, NoH-2, PGL-6, PNN-6, NoH-6}
    \legend{JH-GSR-2, NN-2, J-GSR-2, JH-GSR-6, NN-6, J-GSR-6}
    
\end{axis}
\end{tikzpicture}

%% file: figs/graph_sim.tex
\begin{tikzpicture}[baseline,scale=.93,trim axis left, trim axis right]

\pgfplotstableread{data/graph_sim.csv}\errtable

\begin{semilogyaxis}[
    xlabel={(a) Proportion of different links},
    xmin=.05,
    xmax=.4,
    ylabel={$\sum_{k=1}^K \mathrm{nerr}(\bbSo^{*(k)},\hbSo^{(k)})/K$},
    ymin=.008,
    ymax=1.1,
    grid style=densely dashed,
    grid=both,
    legend style={
        at={(1,0)},
        anchor=south east},
    legend columns=2,
    transpose legend,
    width=185,
    height=160
    ]
    
    \addplot[blue, mark=o, solid] table [x=x, y=PGL-poly] {\errtable}; 
    \addplot[blue!60!white, mark=o, densely dotted] table [x=x, y=PGL-mrf] {\errtable};
    \addplot[red, mark=x, solid] table [x=x, y=LVGL-poly] {\errtable}; \addplot[red!60!white, mark=x, densely dotted] table [x=x, y=LVGL-mrf] {\errtable};
    \addplot[orange, mark=+, solid] table [x=x, y=FGL-poly] {\errtable}; 
    \addplot[orange!60!white, mark=+, densely dotted] table [x=x, y=FGL-mrf] {\errtable}; 

    \legend{JH-GSR-P, JH-GSR-M, LVGL-P, LVGL-M, FGL-P, FGL-M}
    
    % \addplot[blue, mark=o, solid] table [x=x, y=PGL-poly] {\errtable};
    % \addplot[red, mark=x, solid] table [x=x, y=LVGL-poly] {\errtable};
    % \addplot[orange, mark=+, solid] table [x=x, y=FGL-poly] {\errtable}; 
    
    % \addplot[blue!60!white, mark=o, densely dotted] table [x=x, y=PGL-mrf] {\errtable};
    % \addplot[red!60!white, mark=x, densely dotted] table [x=x, y=LVGL-mrf] {\errtable};
    % \addplot[orange!60!white, mark=+, densely dotted] table [x=x, y=FGL-mrf] {\errtable}; 
    
\end{semilogyaxis}
\end{tikzpicture}

%% file: figs/sparsity.tex
\begin{tikzpicture}[baseline,scale=.93,trim axis left, trim axis right]

\pgfplotstableread{data/sparsity01.csv}\errtable

\begin{axis}[
    xlabel={(b) Mean node degree},
    xmin=2,
    xmax=12,
    ylabel={F1-score},
    ymin=0.6,
    ymax=.95,
    grid style=densely dashed,
    grid=both,
    legend style={
        at={(0,0)},
        anchor=south west},
    legend columns=1,
    width=185,
    height=160
    ]
    
    \addplot[blue, mark=o, solid] table [x=xaxis, y=Rwl3_conf1] {\errtable};
    \addplot[red, mark=x, solid] table [x=xaxis, y=Rwl3_conf2] {\errtable};
    
    \addplot[blue!65!white, mark=o, densely dotted] table [x=xaxis, y=Rwl6_conf1] {\errtable};
    \addplot[red!65!white, mark=x, densely dotted] table [x=xaxis, y=Rwl6_conf2] {\errtable};

    \legend{3Rw links ($\alpha=5$ $\beta=100$ $\gamma=300$), 3Rw links ($\alpha=10$ $\beta=50$ $\gamma=200$), 6Rw links ($\alpha=5$ $\beta=100$ $\gamma=300$), 6Rw links ($\alpha=10$ $\beta=50$ $\gamma=200$)}
    % \legend{G1 H-GSR,G1 J-GSR,G1 JH-GSR,G1 J-LVGL,G2 H-GSR,G2 J-GSR,G2 JH-GSR,G2 J-LVGL}
    
\end{axis}
\end{tikzpicture}

%% file: figs/samples_real.tex
\begin{tikzpicture}[baseline,scale=.93,trim axis left, trim axis right]

\pgfplotstableread{data/samples_real.csv}\errtable

\begin{semilogxaxis}[
    % scale only axis,
    xlabel={(a) Number of signals},
    xmin=100,
    xmax=1e6,
    ylabel={$\mathrm{nerr}(\bbSo^{*(k)}, \hbSo^{(k)})$},
    ymin=0,
    ymax=1,
    grid style=densely dashed,
    grid=both,
    legend style={
        at={(1,1)},
        anchor=north east},
    legend columns=2,
    width=185,
    height=160
    ]
    
    \addplot[blue, mark=o, solid] table [x=x, y=G1-J] {\errtable};
    \addplot[blue!65!white, mark=o, densely dotted] table [x=x, y=G1-S] {\errtable};
    \addplot[red, mark=x, solid] table [x=x, y=G2-J] {\errtable};
    \addplot[red!65!white, mark=x, densely dotted] table [x=x, y=G2-S] {\errtable};
    \addplot[orange, mark=+, solid] table [x=x, y=G3-J] {\errtable};
    \addplot[orange!65!white, mark=+, densely dotted] table [x=x, y=G3-S] {\errtable};
    
    % \legend{{G1 Joint}, {G1 Sep}, {G2 Joint}, {G2 Sep}, {G3 Joint}, {G3 Sep} }
    \legend{{G1 JH-GSR}, {G1 SH-GSR}, {G2 JH-GSR}, {G2 SH-GSR}, {G3 JH-GSR}, {G3 SH-GSR} }
    
\end{semilogxaxis}
\end{tikzpicture}

%% file: figs/votes.tex
\begin{tikzpicture}[baseline,scale=.93,trim axis left, trim axis right]

\pgfplotstableread{data/votes.csv}\errtable

\begin{axis}[
    xlabel={(b) Percentage of samples},
    xmin=70,
    xmax=90,
    ylabel={$\mathrm{nerr}(\bbSo^{*(k)}, \hbSo^{(k)})$},
    ymin=0.25,
    ymax=.78,
    grid style=densely dashed,
    grid=both,
    legend style={
        at={(1,1)},
        anchor=north east},
    legend columns=4,
    transpose legend,
    width=185,
    height=160
    ]
    
    \addplot[blue, mark=o, solid] table [x=xaxis, y=G1-GSRH] {\errtable};
    \addplot[red, mark=x, solid] table [x=xaxis, y=G1-JGSR] {\errtable};
    \addplot[purple, mark=*, solid] table [x=xaxis, y=G1-PGL] {\errtable};
    \addplot[orange, mark=+, solid] table [x=xaxis, y=G1-Joint-GL] {\errtable};
    
    \addplot[blue!65!white, mark=o, densely dotted] table [x=xaxis, y=G2-GSRH] {\errtable};
    \addplot[red!65!white, mark=x, densely dotted] table [x=xaxis, y=G2-JGSR] {\errtable};
    \addplot[purple!65!white, mark=*, densely dotted] table [x=xaxis, y=G2-PGL] {\errtable};
    \addplot[orange!65!white, mark=+, densely dotted] table [x=xaxis, y=G2-Joint-GL] {\errtable};

    \legend{G1 SH-GSR,G1 J-GSR,G1 JH-GSR,G1 J-LVGL,G2 SH-GSR,G2 J-GSR,G2 JH-GSR,G2 J-LVGL}
    % \legend{G1 H-GSR,G1 J-GSR,G1 JH-GSR,G1 J-LVGL,G2 H-GSR,G2 J-GSR,G2 JH-GSR,G2 J-LVGL}
    
\end{axis}
\end{tikzpicture}

%% file: figs/votes_4h.tex
\begin{tikzpicture}[baseline,scale=.93,trim axis left, trim axis right]

\pgfplotstableread{data/votes_4h.csv}\errtable

\begin{axis}[
    xlabel={(c) Percentage of samples},
    xmin=70,
    xmax=90,
    ylabel={$\mathrm{nerr}(\bbSo^{*(k)}, \hbSo^{(k)})$},
    ymin=0.25,
    ymax=.78,
    grid style=densely dashed,
    grid=both,
    legend style={
        at={(1,1)},
        anchor=north east},
    legend columns=4,
    transpose legend,
    width=185,
    height=160
    ]
    
    \addplot[blue, mark=o, solid] table [x=xaxis, y=G1-GSRH] {\errtable};
    \addplot[red, mark=x, solid] table [x=xaxis, y=G1-JGSR] {\errtable};
    \addplot[purple, mark=*, solid] table [x=xaxis, y=G1-PGL] {\errtable};
    \addplot[orange, mark=+, solid] table [x=xaxis, y=G1-Joint-GL] {\errtable};
    
    \addplot[blue!65!white, mark=o, densely dotted] table [x=xaxis, y=G2-GSRH] {\errtable};
    \addplot[red!65!white, mark=x, densely dotted] table [x=xaxis, y=G2-JGSR] {\errtable};
    \addplot[purple!65!white, mark=*, densely dotted] table [x=xaxis, y=G2-PGL] {\errtable};
    \addplot[orange!65!white, mark=+, densely dotted] table [x=xaxis, y=G2-Joint-GL] {\errtable};

    \legend{G1 SH-GSR,G1 J-GSR,G1 JH-GSR,G1 J-LVGL,G2 SH-GSR,G2 J-GSR,G2 JH-GSR,G2 J-LVGL}
    % \legend{G1 H-GSR,G1 J-GSR,G1 JH-GSR,G1 J-LVGL,G2 H-GSR,G2 J-GSR,G2 JH-GSR,G2 J-LVGL}
    
\end{axis}
\end{tikzpicture}